\documentclass[useAMS,usenatbib,usegraphicx]{mn2e}

\usepackage{amssymb}
\usepackage{times}
\usepackage{url}
\usepackage{lscape}
\usepackage{booktabs,multirow}
\usepackage{float}
\usepackage{aas_macros} 
\usepackage[usenames]{color}
\newcommand{\kepler}{{\it{Kepler}}}

\title[KIC\,7582608: A new {\it Kepler} roAp star]{KIC\,7582608: A new {\it Kepler} roAp star with frequency variability}

\author[D. L. Holdsworth et al.]
{Daniel L. Holdsworth$^1$\thanks{E-mail:d.l.holdsworth@keele.ac.uk},
B. Smalley$^1$,
D. W. Kurtz$^2$,
J. Southworth$^1$,
\newauthor M. S. Cunha$^3$,
and K. I. Clubb$^4$\\
$^1$Astrophysics Group, Keele University, Staffordshire, ST5 5BG, UK\\
$^2$Jeremiah Horrocks Institute, University of Central Lancashire, Preston, PR1\,2HE, UK\\
$^3$Centro de Astrof\'\i sca e Faculdade de Ci\^encias, Universidade do Porto, Rua das Estrelas 4150-762, Portugal\\
$^4$Department of Astronomy, University of California, Berkeley, CA 94720-3411, USA \\
}
\begin{document}
\date{\today}


\maketitle

\label{firstpage}

\begin{abstract}
We analyse the fifth roAp star reported in the {\it Kepler} field, KIC\,$7582608$, discovered with the SuperWASP project. The object shows a high frequency pulsation at $181.7324$\,d$^{-1}$ (P~$=7.9$\,min) with an amplitude of $1.45$\,mmag, and low frequency rotational modulation corresponding to a period of $20.4339$\,d with an amplitude of $7.64$\,mmag. Spectral analysis confirms the Ap nature of the target, with characteristic lines of Eu\,{\sc{ii}}, Nd\,{\sc{iii}} and Pr\,{\sc{iii}} present. The spectra are not greatly affected by broadening, which is consistent with the long rotational period found from photometry. From our spectral observations we derive a lower limit on the mean magnetic field modulus of $\langle B \rangle = 3.05 \pm 0.23$\,kG. Long Cadence {\it Kepler} observations show a frequency quintuplet split by the rotational period of the star, typical for an oblique pulsator. We suggest the star is a quadrupole pulsator with a geometry such that $i\sim66^\circ$ and $\beta\sim33^\circ$. We detect frequency variations of the pulsation in both the WASP and {\it Kepler} data sets on many time scales. Linear, non-adiabatic stability modelling allows us to constrain a region on the HR diagram where the pulsations are unstable, an area consistent with observations.
\end{abstract}

\begin{keywords}
asteroseismology -- stars: chemically peculiar -- stars: oscillations -- stars: individual: KIC\,7582608 -- techniques: photometric.
\end{keywords}

\section{Introduction}
\label{sec:intro}

The rapidly oscillating Ap (roAp) stars can be found at the base of the classical instability strip on the Hertzsprung-Russell (HR) diagram, between the zero-age main-sequence and terminal-age main-sequence in luminosity. Since their first observation over $30$\,years ago \citep{kurtz82}, about $60$ stars of this type have been discovered. Most stars in this class have been discovered with ground-based photometric observations of known Ap stars. However, there are now some stars that have been detected through spectral line variations alone \citep[e.g.][]{elkin10,elkin11,kochukhov13} and, with the launch of the {\it Kepler} satellite, high-precision space-based photometry \citep[e.g.][]{balona11a,balona11b,kurtz11,niemczura14}.

The roAp stars show pulsational periods in the range of $6$ to $23$\,min with amplitudes in Johnson $B$ of up to $10$\,mmag. These stars have their pulsational and magnetic axes inclined to the rotational axis, leading to the oblique pulsator model (\citealt{kurtz82}; \citealt{bigot02}; \citealt{bigot11}). Such an orientation allows the pulsation modes to be viewed from different aspects throughout the stellar rotational cycle.

The observations of roAp stars provide the best laboratory, beyond the Sun, to study the interactions between strong global magnetic fields, between $1-24$\,kG, and stellar pulsations. Although pulsating in roughly the same frequency range, the driving mechanisms differ between the Sun and the roAp stars: the $\kappa$-mechanism acting on the hydrogen ionisation zone is thought to be the driving force of the roAp high-overtone pressure mode (p-mode) pulsations, although recent models by \citet{cunha13} suggest another excitation mechanism may be at work for a subset of roAp stars, whereas the solar pulsations are stochastically driven in the convection zone.

The Ap stars typically show strong over-abundances of rare-earth elements in their atmospheres of up to $10^6$ times the solar value. Furthermore, these chemical anomalies often manifest themselves as spots on the stellar surface caused by atomic diffusion and trapping of ions by the strong magnetic field. These surface brightness anomalies are stable over many decades, allowing the rotation period of the star to be well determined.

To date there have been no roAp stars found in close binary systems. In fact, few Ap stars are in close binaries. There are, however, three candidates which are strongly suspected of being in visual binary systems (i.e. HR\,3831, $\gamma$\,Equ and $\alpha$\,Cir). \citet{scholler12} conducted a study of known roAp stars in the search for companions using near-infrared imaging. They found six of their target roAp stars showed signs of companions, two of which were already known. However, from photometric data alone it is not possible to categorically determine whether the companions and the targets are gravitationally bound. Instead they calculate the probability of a chance projection, with their least certain companion having a chance projection probability of less than $2\times10^{-3}$. Regardless of whether they are gravitationally bound, the separation of visual binary stars means that stellar physics governing the individual stars is not effected. It is only in close binary systems that the components influence each other.

As previously mentioned, the {\it Kepler} mission has enabled the discovery of four roAp stars, all which have pulsation amplitudes much below the detection limits of ground-based photometry. The first to be found in the {\it Kepler} data was KIC\,8677585 \citep{balona11a}, a known A5p star observed during the 10\,d commissioning run with the 1\,min short cadence (SC) mode. This roAp star pulsates at multiple frequencies in the range $125-145$\,d$^{-1}$ with amplitudes in the range of $8.4-32.9$\,$\mu$mag. This star also shows a low frequency variation at about 3\,d$^{-1}$ which \citeauthor{balona11a} suggest is a g-mode $\gamma$ Doradus pulsation, after consideration of other possibilities.

The second roAp star in the {\it Kepler} data is KIC\,10483436 \citep{balona11b} pulsating at two frequencies, each mode showing a quintuplet separated by the well determined rotation period. Again, the amplitudes of pulsation are much below the ground-based detection limits, namely in the range $5-69$\,$\mu$mag.

KIC\,10195926 is the third identified roAp star in {\it Kepler} data \citep{kurtz11}. As with KIC\,10483436, KIC\,10195926 pulsates in two independent modes with amplitudes less than $170$\,$\mu$mag. Both modes show rotational splitting. The precision of the {\it Kepler} data has allowed the authors to claim KIC\,10195926 has two pulsational axes, the first evidence of such a phenomenon in roAp stars.

The forth {\it Kepler} roAp is KIC\,4768731 \citep{niemczura14}. Discovered through spectral classification, KIC\,4768731 is the second slowest roAp pulsator, varying with a frequency of 61.45\,d$^{-1}$ at an amplitude of $62.6$\,$\mu$mag (Smalley et al., in preparation).

The fifth roAp star found in the {\it Kepler} field, which we analyse here, is KIC\,7582608. Discovered using data from the SuperWASP project \citep[][their `J1844']{holdsworth14}, KIC\,7582608 shows a single pulsation at $181.7324$\,d$^{-1}$ with an amplitude of $1.45$\,mmag in the WASP broad-band filter ($4000-7000$\,\AA). The star was observed by the {\it Kepler} satellite for the full duration of the mission in the 30\,min, Long Cadence (LC) mode. As a result, the true pulsation frequency is above the Nyquist frequency of the {\it Kepler} LC data. Although this results in a reduction of the observed pulsation amplitude, it has been shown that frequencies higher than the Nyquist can be reliably extracted from LC data \citep{murphy13}. 

We present an analysis of spectral observations of KIC\,7582608 which allows us to place the star on the theoretical HR diagram. We then show the WASP data used to initially detect the pulsations, and an in-depth study of the object with the {\it Kepler} data, followed by the results of linear, non-adiabatic modelling of the star.

\section{Spectroscopic observations}
\label{sec:spec}

We have obtained two epochs of spectra for KIC\,7582608 using the Hamilton Echelle Spectrograph (HamSpec) mounted on the $3.0$-m Shane telescope at Lick Observatory \citep{vogt87}. The observations were taken on 2012 July 24 and 2013 June 28 and have a resolution of $R\sim37\,000$ and a signal-to-noise ratio (S/N) of about $40$. The spectra were reduced in IDL with instrument specific software that performs flat-field corrections, de-biases, cosmic-ray cleaning, and wavelength calibrations. We co-added three spectra from 2012 and two from 2013 to produce two single spectra. Table\,\ref{tab:spec} gives the spectral details, including the rotation phase at which the spectra were taken (see Section\,\ref{sec:wasp-rot}).

\begin{table}
  \centering
  \caption{Details of spectroscopic observations taken with HamSpec on the Shane $3.0$-m telescope. The rotational phase has been calculated as shown in the text.}
  \label{tab:spec}
    \begin{tabular}{ccc}
      \hline
      \multicolumn{1}{c}{BJD-245\,0000.0} &
      \multicolumn{1}{c}{Exposure time} &
      \multicolumn{1}{c}{Rotational phase}\\
       & (s) &  \\
      \hline
      6132.81003 & 1800 & 0.88 \\
      6132.83394 & 1800 & 0.88 \\
      6132.85516 & 1800 & 0.88 \\

      6471.79915 & 1800 & 0.47 \\
      6471.82036 & 1800 & 0.47 \\
      \hline
      \end{tabular}
\end{table}

The two spectra show velocity shifts away from rest wavelengths for each epoch. These shifts are calculated by comparison with strong, unblended lines of Mg\,{\sc{ii}} $4481$\,\AA, Ca\,{\sc{ii}} $4226$\,\AA\,and the Na\,D lines, resulting in shifts of $45.8\pm2.8$ and $50.5\pm2.8$\,km\,s$^{-1}$ for 2012 and 2013, respectively.

\begin{figure}
  \centering
  \includegraphics[angle=180,width=27mm]{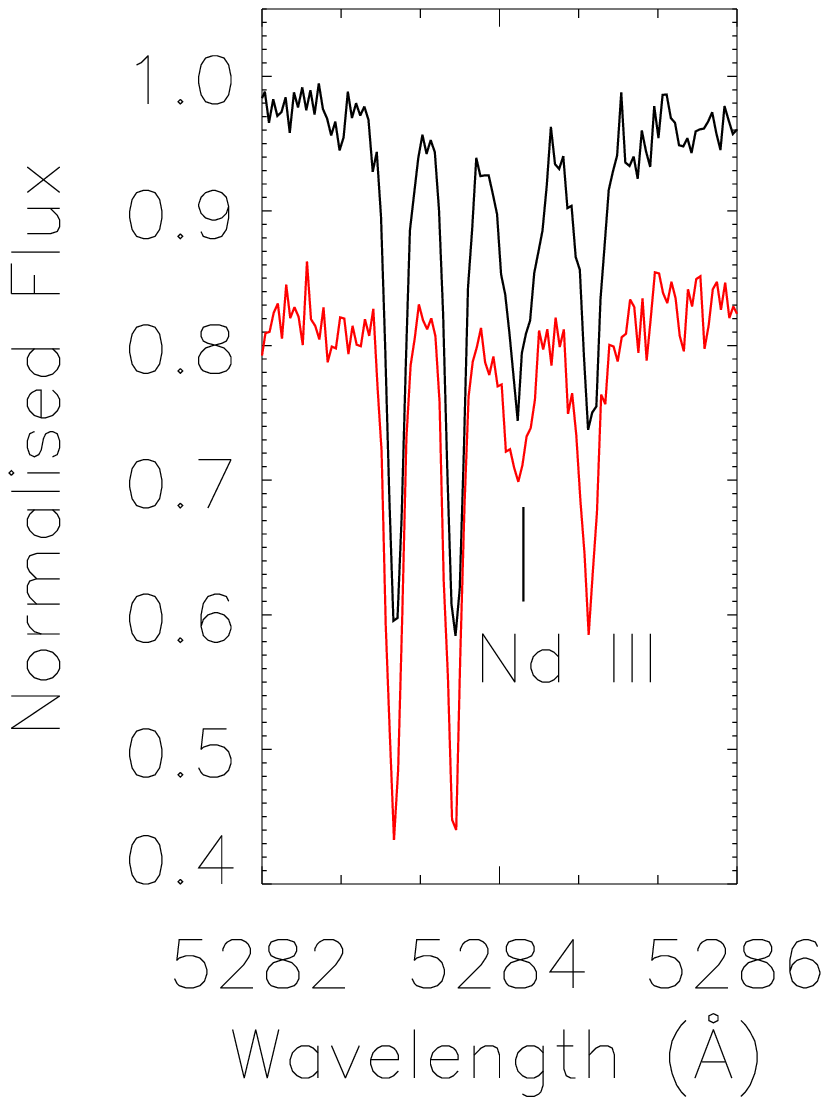}
  \includegraphics[angle=180,width=27mm]{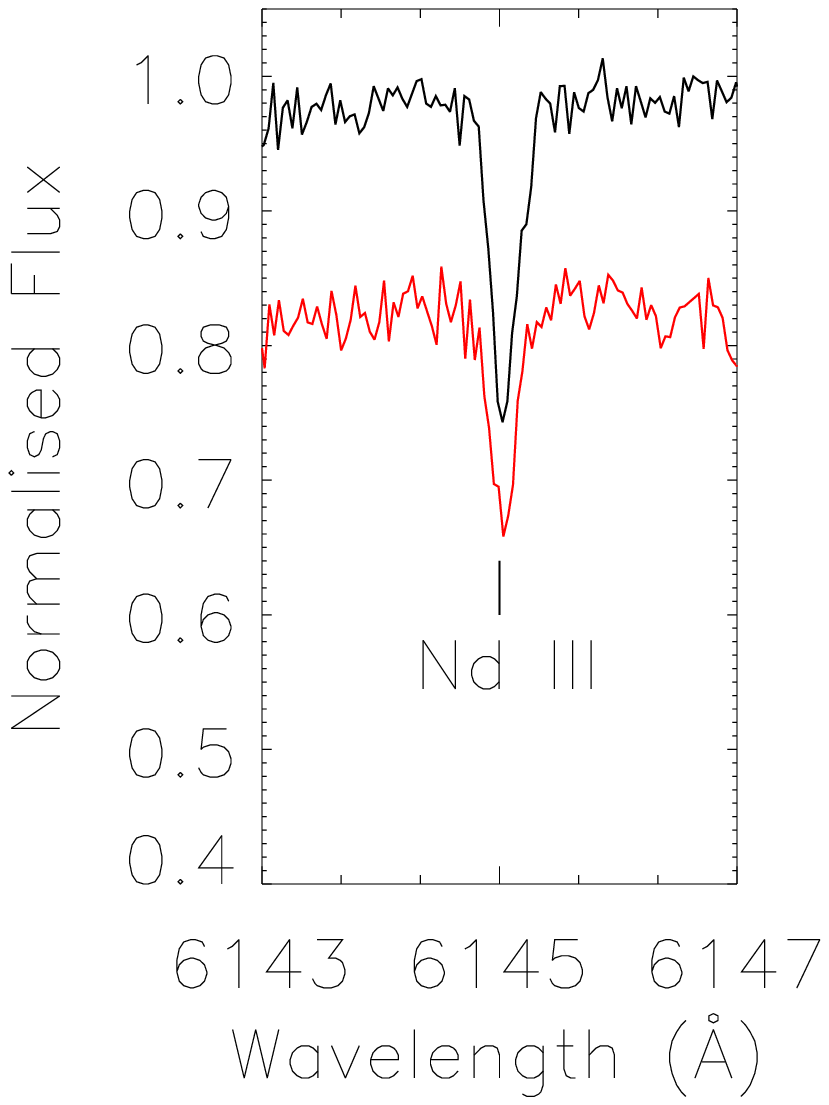}
  \includegraphics[angle=180,width=27mm]{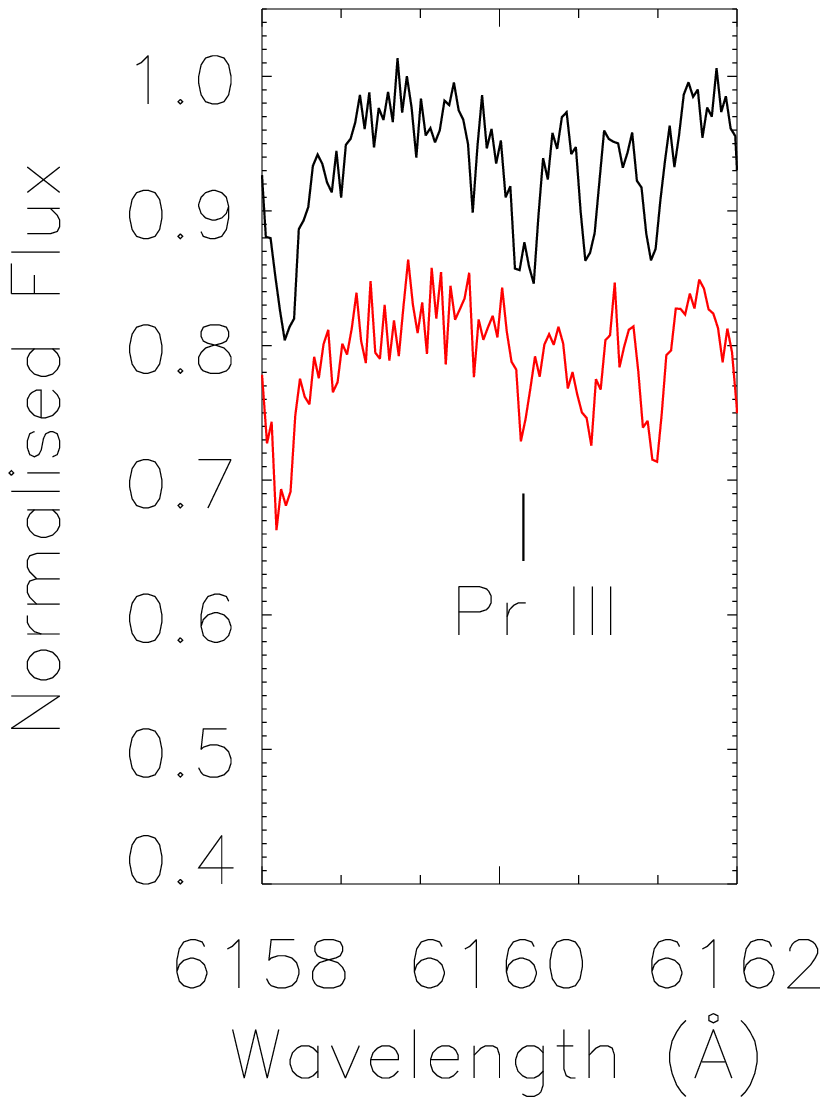}

  \includegraphics[angle=180,width=27mm]{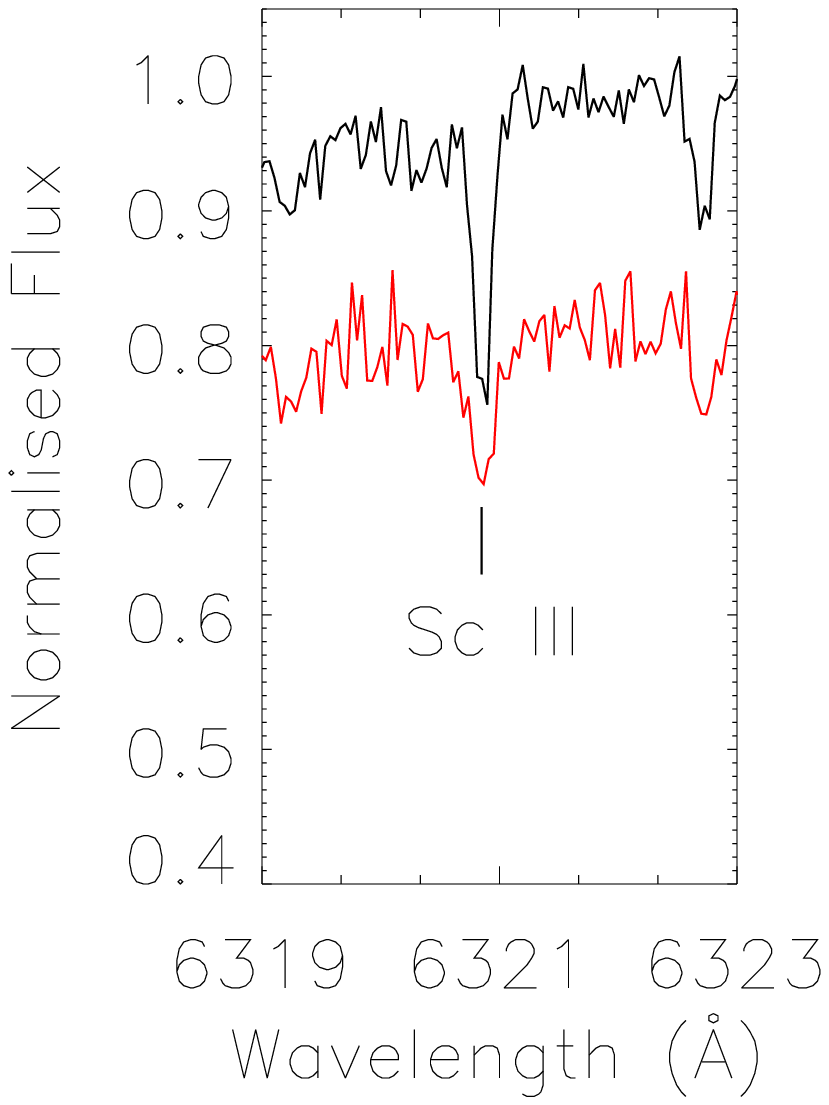}
  \includegraphics[angle=180,width=27mm]{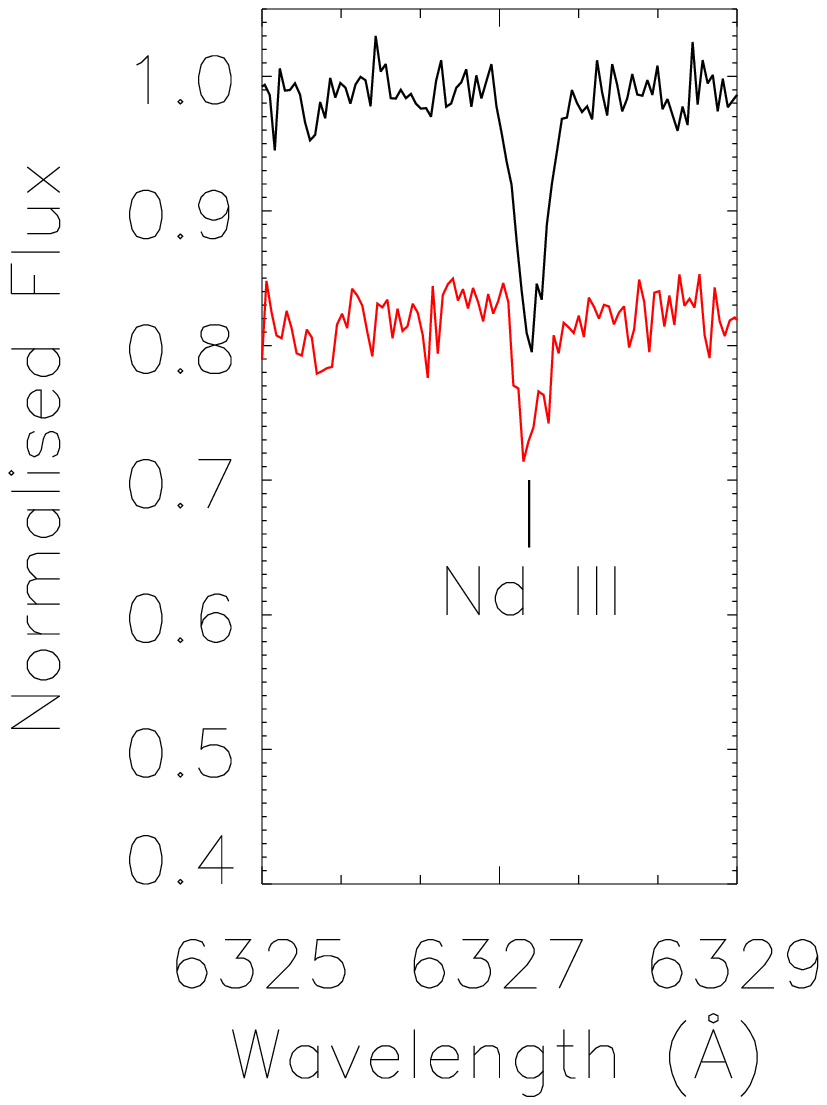}
  \includegraphics[angle=180,width=27mm]{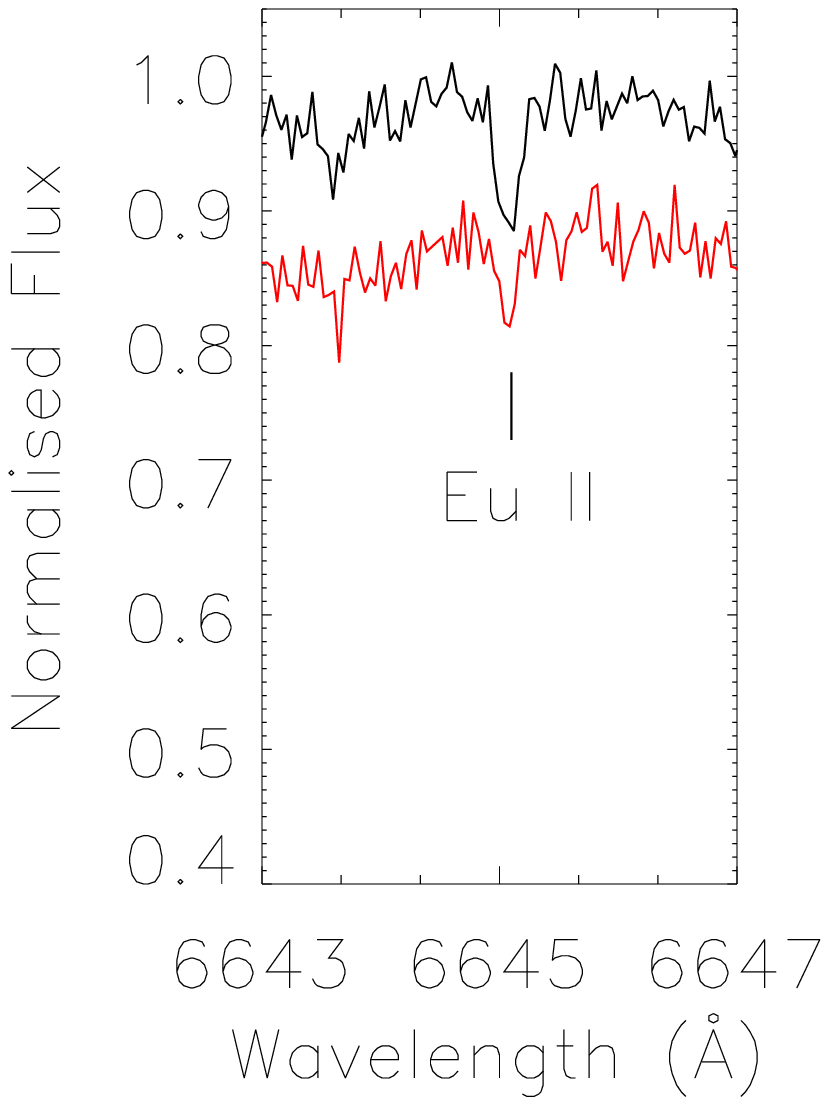}

\caption{A selection of spectral lines which demonstrate the Ap nature of KIC\,7582608. The upper black line shows the 2012 spectrum, with the lower red line showing the 2013 spectrum offset by 0.15 for clarity. There is clear variability in the strength of the peculiar lines due to the rotational phase variability. Both spectra have had velocity shifts removed to align the spectral lines with their rest wavelengths.}
\label{fig:spec}
\end{figure}

Figure\,\ref{fig:spec} gives examples of spectral lines of elements found in the atmospheres of Ap stars extracted from our spectra. We note the variation of line strengths between the two different epochs and thus rotational phase. Close to maximum optical light, at phase $0.88$ in 2012 (the upper black lines in Fig.\,\ref{fig:spec}), we see stronger lines of the rare earth elements compared to the second epoch at phase $0.47$ (red lower lines in Fig.\,\ref{fig:spec}) where we observe the opposite hemisphere. The ephemeris for the phase ($\phi$) calculations are calculated for the first light maximum (BJD) in the WASP data such that
\begin{equation}
\phi(E)=245\,3151.8797+\mbox{20{\fd}4339}\times E
\end{equation}
where $E$ is the number of cycles elapsed since the initial epoch.

The spectra show sharp lines indicative of a low $v\sin i$. We are unable to determine an accurate value of the $v\sin i$ from these spectra as we are limited by the spectral resolution, but we estimate $v \sin i \le 4$\,km\,s$^{-1}$. High resolution spectra are required to constrain this value.

The spectral lines typically used to determine the mean magnetic field modulus, $\langle B \rangle$, such as Cr\,{\sc{ii}} $6147.7$\,\AA\ and Fe\,{\sc{ii}} $6147.7$ and $6149.2$\,\AA\ \citep{mathys90}, show no sign of Zeeman splitting; however we attribute this to a lack of resolution in our spectra. Instead we use the method of \citet{mathys92} to estimate this value by comparing the relative intensities of the Fe\,{\sc{ii}} lines at $6147.7$ and $6149.2$\,\AA. There is a change of relative strengths between the two spectra, however the $6147.7$ line remains the stronger of the two. We determine $\langle B \rangle$ = $3.13 \pm 0.32$ and $2.97 \pm 0.32$ kG for the 2012 and 2013 spectra, respectively. It is noted by \citet{mathys92} that for their stars with no resolvable magnetically split lines, this method of comparing Fe {\sc{ii}} $6147.7$ and $6149.2$\,\AA\ produces systematically lower values of the mean magnetic field modulus. We therefore suggest that the values derived here for KIC\,7582608 are representative of a lower limit on $\langle B \rangle$.

We used {\sc{uclsyn}} \citep{smith88,smith92,uclsyn} to determine the $\log g$ and $T_{\rm eff}$ by performing an abundance analysis by measuring the equivalent widths of 77 Fe\,{\sc{i}} lines and 57 Fe\,{\sc{ii}} lines. The effective temperature was found to be $T_{\rm eff} = 8700 \pm 100$\,K by requiring no dependence of abundance with excitation potential. For the $\log g$ determination, we tested the ionisation balance between Fe\,{\sc{i}} and Fe\,{\sc{ii}}. We found $\log g = 4.3\pm0.4$ (cgs). The photometrically derived value presented in the {\it Kepler} Input Catalogue (KIC) of 4.1 agrees with our calculation. These results were also confirmed by minimising the scatter in the measured abundances.

The Fe abundance measured from the 77 lines is $\log {\rm A(Fe)} =8.4 \pm 0.2$, indicating an over-abundance relative to solar of ${\rm [Fe/H]} = +0.9$\,dex based on the solar chemical composition presented by \citet{asplund09}.

We have also determined the $T_{\rm eff}$ of KIC\,7582608 using stellar spectral energy distribution (SED) fitting. We have used literature photometry from 2MASS \citep{skrutskie06}, $U$, $B$ and $V$ magnitudes from \citet{everett12}, $g'$, $r'$ and $i'$ from \citet{greiss12}, the TASS $I$ magnitude  \citep{droege06}, and GALEX fluxes \citep{martin05} to reconstruct the SED. To determine the reddening of the star, we measured the Na\,D lines from our spectra and applied the relation of \citet{munari97} to derive an $E(B-V) = 0.04 \pm 0.02$.

The stellar $T_{\rm eff}$ value was determined by fitting a [M/H]$\,= \,0.0$  \citet{kurucz93} model to the de-reddened SED. The model fluxes were convolved with photometric filter response functions. A weighted Levenberg-Marquardt non-linear least-squares fitting procedure was used to find the solution that minimised the difference between the observed and model fluxes. We used the $\log g = 4.3 \pm 0.4$ derived from the spectra for the fit. The uncertainty in $T_{\rm eff}$ includes the formal least-squares error and adopted uncertainties in $E(B-V)$ of $\pm0.02$, $\log g$ of $\pm0.4$ and [M/H] of $\pm0.5$ added in quadrature. As a result of the SED fitting, we derive a temperature of $8670 \pm 450$\,K for KIC\,7582608. Assuming a standard error of $200$\,K for the KIC temperature, their quoted value of $8149\pm200$\,K agrees with our measurements. 

\section{Position in the HR Diagram}
\label{sec:HR}

To place KIC\,7582608 in a theoretical HR diagram, we used both our calculated values from the spectra and adopted the values from the KIC. From the KIC, the values are $R = 1.82$\,R$_{\odot}$, $T_{\rm eff} = 8149$\,K, and $\log g = 4.10$. From these we calculate the star's luminosity, $L$, as $\log L/ L_{\odot}=1.12$.

Using the values derived from the spectra, namely $T_{\rm eff}=8700\pm100$\,K, $\log g=4.3\pm0.4$ and [Fe/H]$=0.9\pm0.5$, and the calibrations from \citet{torres10}, we derive the mass, radius and luminosity of KIC\,7582608 of: $M=2.37\pm0.43$\,M$_\odot$, $R=1.77\pm0.92$\,R$_\odot$ and $\log L/L_\odot=1.21\pm0.45$.

\begin{figure}
  \includegraphics[width=\linewidth,angle=180]{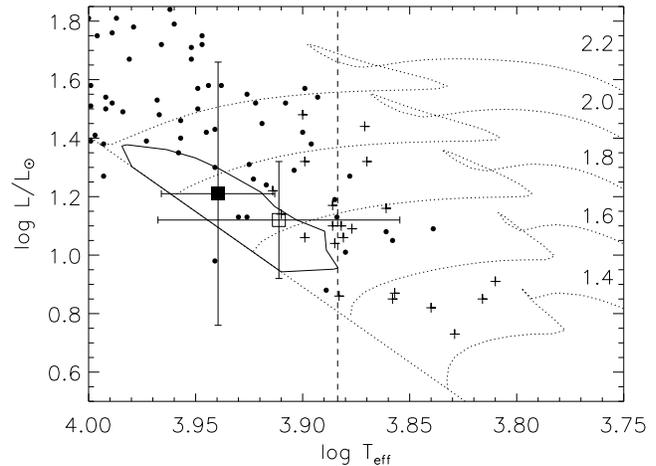}
  \caption{The position of KIC 7582608 in the theoretical HR diagram. The filled square is calculated with the temperature from the spectra and the derived radius, while the open square is calculated from the KIC values. The other roAp stars (crosses) and noAp stars (dots) are also shown for context. The vertical dashed line indicates the lower temperature tested for the models, and the area enclosed by the solid line represents the region where unstable modes are present at the observed frequency of $181.7324$\,d$^{-1}$ (see Section\,\ref{sec:model} for further details). The zero-age main-sequence and evolutionary tracks are from \citet{bertelli08}.}
  \label{fig:HR}
 \end{figure}

The positions of KIC\,7582608 are shown in Fig.\,\ref{fig:HR} along with other roAp stars and non-oscillating Ap (noAp) stars for which temperatures and luminosities are available in the literature. The position of the open square is determined from the KIC values, whilst the filled square is from spectral values. In both cases, KIC\,7582608 appears to be close to the zero-age main-sequence, and is amongst the hotter, if not the hottest, known roAp stars. Further spectral observations are required to improve the effective temperature and luminosity measurements.

It is important to note here that both positions fall within the area bound by the solid lines. It is in this region that modelling has shown unstable pulsation modes of the same frequency as that observed in KIC\,7582608. The modelling will be addressed in full in Section\,\ref{sec:model}.

\section{SuperWASP discovery data}
\label{sec:wasp}

We present here the data which led to the identification of KIC\,7582608 as a roAp star. We provide a short summary of the WASP instrument and data here and refer the reader to \citet{pollacco06} and \citet{holdsworth14} for details of the WASP project and the techniques used to identify KIC\,7582608, respectively.

The SuperWASP instruments consist of eight $200$\,mm, f$/1.8$\,Canon telephoto lenses backed by Andor CCDs of $2048\times2048$\,pixels observing $\sim$61\,deg$^{2}$\,each through broadband filters covering a wavelength range of $4000-7000$\,\AA. This set-up enables simultaneous observations of up to $8$ fields with a pixel size of $13.7$\,arcsec. The instruments capture two consecutive $30$\,s integrations at a given pointing, then move to the next observable field. Typically, fields are revisited every $10$\,min. The data pass through a reduction pipeline correcting for primary and secondary extinction, the colour response of the instrument and the zero-point. The data are also corrected for instrumental systematics with the SysRem algorithm of \citet{tamuz05}. Individual stellar magnitudes are extracted based on positions from the USNO-B1.0 catalogue \citep{monet03} down to a limit of about magnitude 15.

KIC\,7582608 was initially observed by WASP in 2004, and subsequently in 2007, 2008, 2009 and 2010, denoted as `seasons' (there are multiple observing blocks per year if the target appears in more than one observing field, in this case denoted a and b in Table\,\ref{tab:wasp-obs}). Observations are taken in a variety of conditions that can result in significant errors and night to night fluctuations in the final data. Table\,\ref{tab:wasp-obs} details the WASP observations of KIC\,7582608, with the final column representing the weighted reduced-$\chi^2$ \citep{bevington69} which aims to characterise the data using the number of points and scatter in the light curve such that

\begin{equation}
\chi^2/n=\displaystyle{\frac{\Sigma(({\rm{mag}}-\rm{median}({\rm{mag}}))/\sigma)^2}{(n-1)}}.
\end{equation}

\begin{table}
  \centering
  \caption{Details of WASP observations of KIC\,7582608. BJD is given as BJD$-$254\,0000.0}
  \label{tab:wasp-obs}
  \begin{tabular}{ccccc}
    \hline
    \multicolumn{1}{c}{WASP}&
    \multicolumn{1}{c}{BJD}&
    \multicolumn{1}{c}{Length}&
    \multicolumn{1}{c}{Number of}&
    \multicolumn{1}{c}{$\chi^{2}/n$}\\

    \multicolumn{1}{c}{season}&
    \multicolumn{1}{c}{start}&
    \multicolumn{1}{c}{(d)}&
    \multicolumn{1}{c}{data points}&
    \multicolumn{1}{c}{}\\

    \hline

    \multicolumn{1}{c}{2004a}&
    \multicolumn{1}{c}{3139.67092}&
    \multicolumn{1}{c}{138.69311}&
    \multicolumn{1}{c}{$1967$}&
    \multicolumn{1}{c}{$3.47$}\\

    \multicolumn{1}{c}{2004b}&
    \multicolumn{1}{c}{3139.66262}&
    \multicolumn{1}{c}{116.82519}&
    \multicolumn{1}{c}{$1586$}&
    \multicolumn{1}{c}{$4.15$}\\

    \multicolumn{1}{c}{2004$^*$}&
    \multicolumn{1}{c}{3139.66262}&
    \multicolumn{1}{c}{138.70141}&
    \multicolumn{1}{c}{$3553$}&
    \multicolumn{1}{c}{$3.78$}\\

    \multicolumn{1}{c}{2007a}&
    \multicolumn{1}{c}{4230.56374}&
    \multicolumn{1}{c}{66.10058}&
    \multicolumn{1}{c}{$3381$}&
    \multicolumn{1}{c}{$0.84$}\\

    \multicolumn{1}{c}{2007b}&
    \multicolumn{1}{c}{4297.39089}&
    \multicolumn{1}{c}{38.08300}&
    \multicolumn{1}{c}{$1660$}&
    \multicolumn{1}{c}{$1.16$}\\

    \multicolumn{1}{c}{2007$^*$}&
    \multicolumn{1}{c}{4230.56374}&
    \multicolumn{1}{c}{104.91015}&
    \multicolumn{1}{c}{$5041$}&
    \multicolumn{1}{c}{$0.94$}\\

    \multicolumn{1}{c}{2008}&
    \multicolumn{1}{c}{4577.59791}&
    \multicolumn{1}{c}{112.95654}&
    \multicolumn{1}{c}{$7585$}&
    \multicolumn{1}{c}{$10.04$}\\

    \multicolumn{1}{c}{2009}&
    \multicolumn{1}{c}{4941.57840}&
    \multicolumn{1}{c}{125.93750}&
    \multicolumn{1}{c}{$10917$}&
    \multicolumn{1}{c}{$1.08$}\\

    \multicolumn{1}{c}{2010}&
    \multicolumn{1}{c}{5307.57254}&
    \multicolumn{1}{c}{124.95556}&
    \multicolumn{1}{c}{$12293$}&
    \multicolumn{1}{c}{$1.00$}\\

    \hline
    \multicolumn{5}{l}{$^*$Combined data sets a and b.}\\
  \end{tabular}
\end{table}

As can be seen from Table\,\ref{tab:wasp-obs}, the 2004 and, especially, the 2008 seasons of data have a $\chi^2/n$ value deviating from the desired value of 1.00 indicating that these seasons may result in less reliable results.

\subsection{The rotation signature}
\label{sec:wasp-rot}

\begin{figure}
  \includegraphics[angle=180,width=\linewidth]{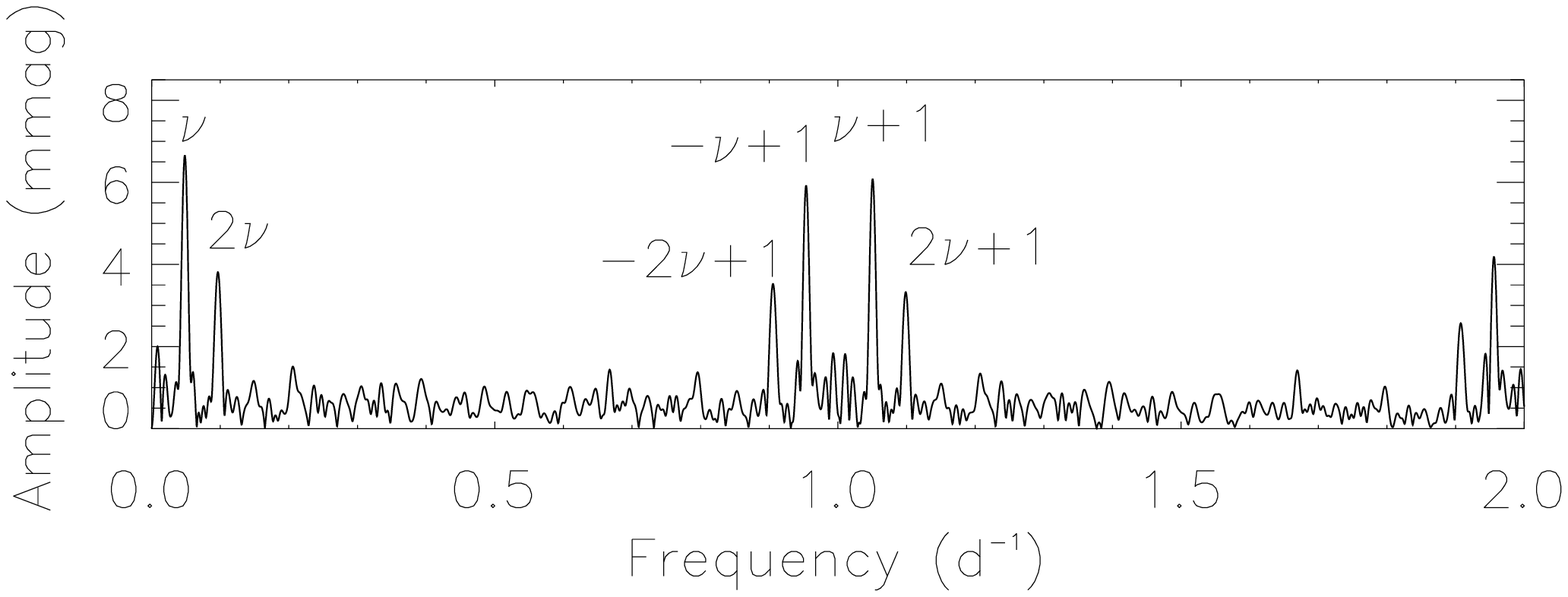}\hfill
  \includegraphics[angle=180,width=\linewidth]{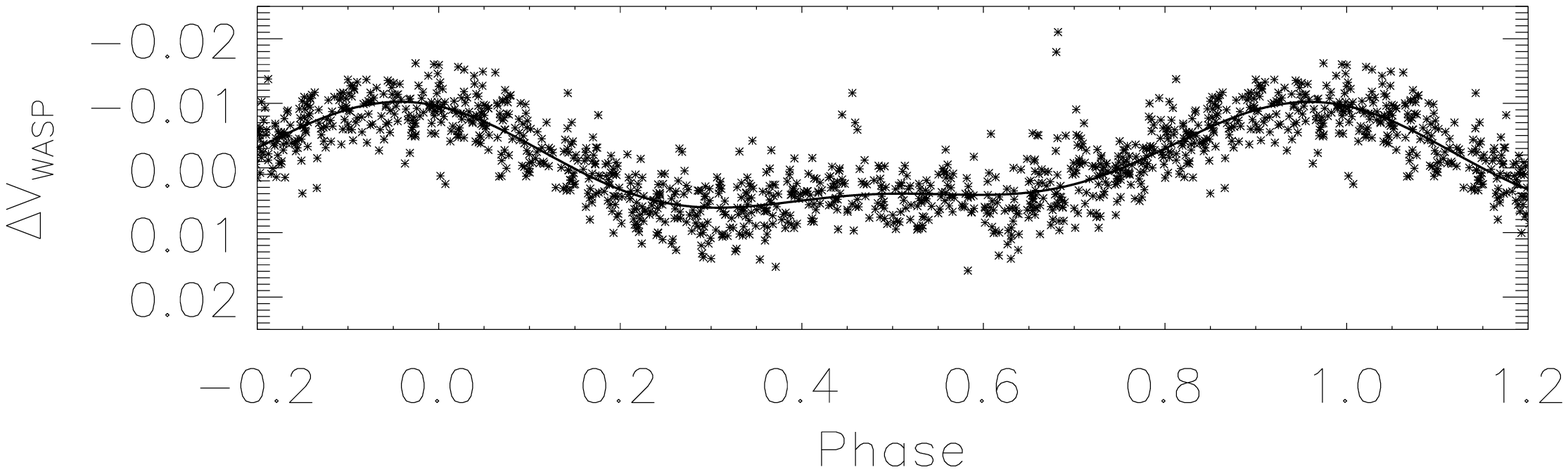}
  \caption{Top panel: labelled periodogram of the 2010 season of WASP data, where $\nu_{\rm rot}$ is the rotation frequency. Bottom panel: the phase folded light curve of the 2010 WASP data. The data are folded on the period derived from combining the 2007, 2009 and 2010 seasons, i.e. $20.4339$\,d and are shown in phase bins of $10:1$.}
  \label{fig:wasp-rot}
\end{figure}

The WASP light curve for KIC\,7582608 shows the strong modulation which is typical of Ap stars. We calculated a Lomb-Scargle periodogram using the Fortran code {\sc{fasper}} \citep{press89,press92} in the frequency range $0-2$\,d$^{-1}$. We used the best quality data for the calculation spanning $1201.9644$\,d to best constrain the period, i.e. the 2007, 2009 and 2010 seasons. Using P{\sc{eriod}}04 \citep{lenz05}, we found a rotation frequency of $\nu_{\rm rot} = 0.0489\,382 \pm 0.000\,004$\,d$^{-1}$, which corresponds to a rotation period of $P_{\rm rot}=20.4339\pm0.0017$\,d. The errors are the analytical errors taken from P{\sc{eriod}}04 \citep{montgomery99}. Fig.\,\ref{fig:wasp-rot} shows a labelled periodogram and phase folded light curve of just the 2010 season for clarity. As well as the rotation frequency we detect a second weaker signature at $2\nu_{\rm rot}$. In the bottom panel of Fig.\,\ref{fig:wasp-rot} this can be seen as a small bump at phase 0.5. This signature can be explained with another spot on the opposite hemisphere of the star which appears smaller, most likely due to projection effects.

\subsection{The pulsation signature}
\label{sec:WASP-puls}

\begin{figure}
  \includegraphics[angle=180,width=\linewidth]{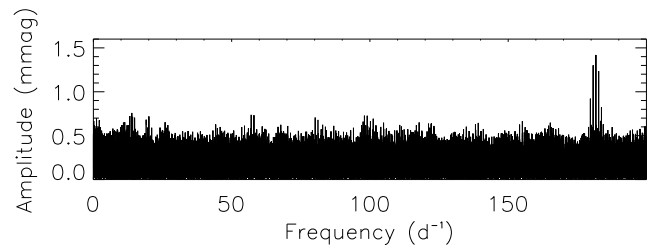}
  \caption{Periodogram of the 2010 season WASP  data after pre-whitening has occurred. The pulsation is clearly seen at $181.7324$\,d$^{-1}$ at an amplitude of about 1.45\,mmag.}
  \label{fig:WASP_puls}
\end{figure}

To best analyse the pulsation in the WASP data, we pre-whitened the data to $10$\,d$^{-1}$ by fitting a series of sinusoids to a limit of the approximate noise level at high-frequency for each data set individually. The noise level varied greatly between seasons, with pre-whitening occurring between 0.7 and 2.5\,mmag for the flattest and most noisy data sets, respectively. In this way we removed the rotational variability from the light curve, and any further systematic effects, especially the dominant `red' noise \citep{smith06} remaining after the data have passed through the WASP pipeline. The resulting periodogram for the 2010 season is shown in Fig.\,\ref{fig:WASP_puls}. The results of a non-linear least-squares fit is shown for each season in Table\,\ref{tab:wasp-nls}.

\begin{table}
  \centering
    \caption{Frequencies, amplitudes and phases of the pulsation extracted from the WASP data using a non-linear least-squares fit. The zero-point for the phases is BJD 245\,3151.6245.}
    \label{tab:wasp-nls}
    \begin{tabular}{cccc}
      \hline
      \multicolumn{1}{c}{Season}&
      \multicolumn{1}{c}{Frequency}&
      \multicolumn{1}{c}{Amplitude}&
      \multicolumn{1}{c}{Phase}\\
      
      \multicolumn{1}{c}{}&
      \multicolumn{1}{c}{(d$^{-1}$)}&
      \multicolumn{1}{c}{(mmag)}&
      \multicolumn{1}{c}{(rad)}\\

      \hline

      2004a &	$181.7247\pm0.0011$   &    $1.429\pm0.337$  &     $ 1.644\pm0.238$ \\  
      2004b &   $181.7243\pm0.0017$   &    $1.444\pm0.435$  &     $ 1.865\pm0.307$ \\  
      2004  &   $181.7241\pm0.0013$   &    $1.274\pm0.316$  &     $ 1.827\pm0.252$ \\  
      2007a &   $181.7339\pm0.0008$   &    $1.452\pm0.222$  &     $-1.386\pm0.152$ \\  
      2007b &   $181.7353\pm0.0009$   &    $1.638\pm0.373$  &     $ 1.577\pm0.228$ \\  
      2007  &   $181.7347\pm0.0006$   &    $1.549\pm0.195$  &     $-0.769\pm0.125$ \\  
      2008  &   $181.7334\pm0.0011$   &    $1.597\pm0.349$  &     $ 1.465\pm0.220$ \\  
      2009  &   $181.7387\pm0.0004$   &    $1.460\pm0.154$  &     $ 0.724\pm0.105$ \\  
      2010  &   $181.7276\pm0.0005$   &    $1.436\pm0.158$  &     $ 2.067\pm0.109$ \\  

      \hline
    \end{tabular}
\end{table}

Other than extracting the correct pulsational frequency and amplitude (in the WASP photometric pass-band), there is no further information that can be gleaned from the WASP data. As previously mentioned, the pulsation frequency is above the Nyquist frequency of {\it Kepler} LC data and some confusion may be had in disentangling the true peak from the many other aliases if the data do not have a large enough time span, a scenario which the WASP data allows us to disregard. The noise characteristics of the WASP data do not allow for a reliable extraction of rotational sidelobes as they occur at approximately the same amplitude. We do however notice that the frequencies of the different seasons vary beyond their errors, an observation we shall revisit in Section\,\ref{sec:freqvar}.

\section{{\it Kepler} observations}
\label{sec:kplr}

As previously mentioned, KIC\,7582608 has been observed by the {\it Kepler} satellite for the full duration of the mission, a little over 4\,y. The data were collected in the LC mode with a cadence of 30\,min. There are no SC data for this target.

\subsection{The  rotation signature}
\label{sec:kplr-rot}

\begin{figure}
  \includegraphics[angle=180,width=\linewidth]{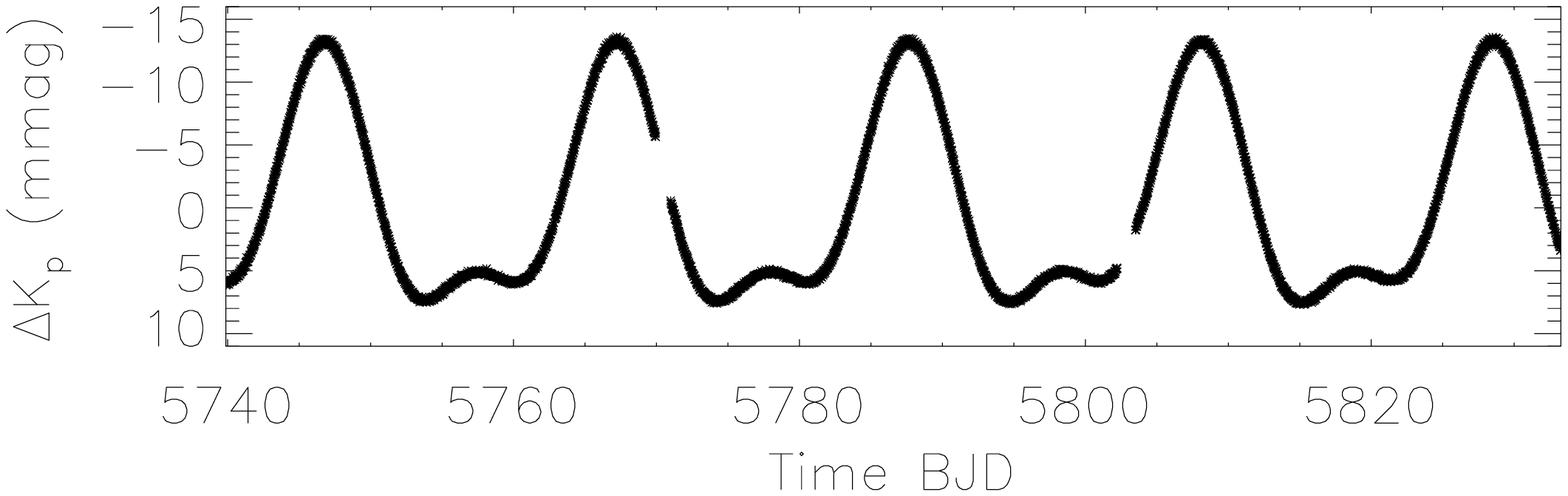}
  \includegraphics[angle=180,width=\linewidth]{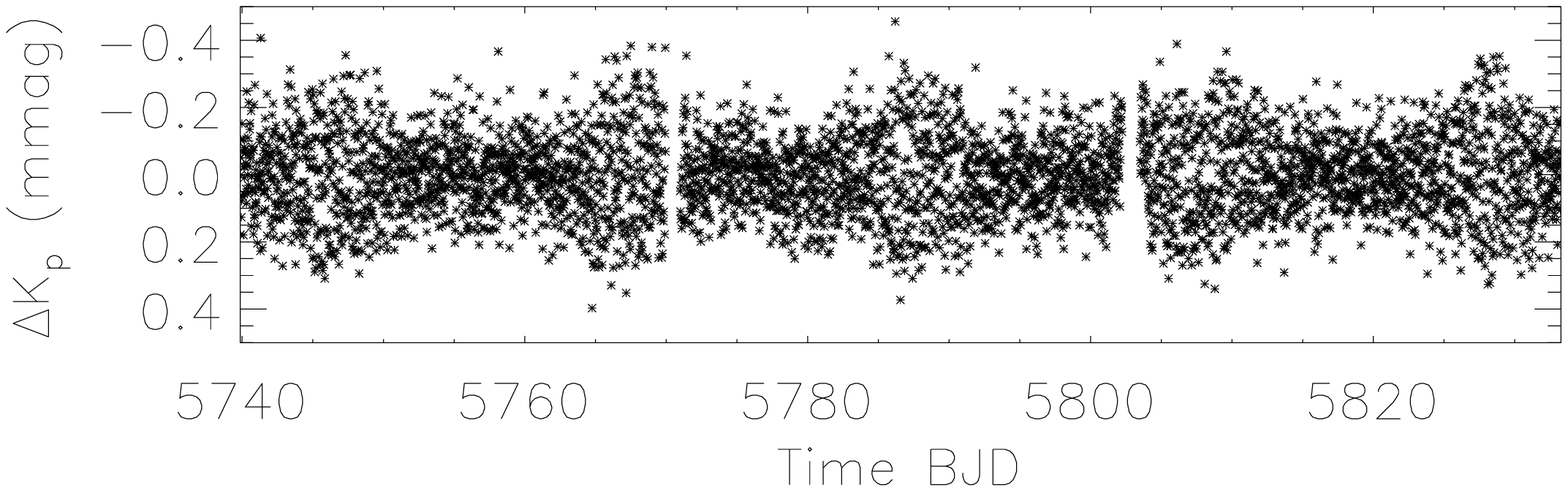}
  \includegraphics[angle=180,width=\linewidth]{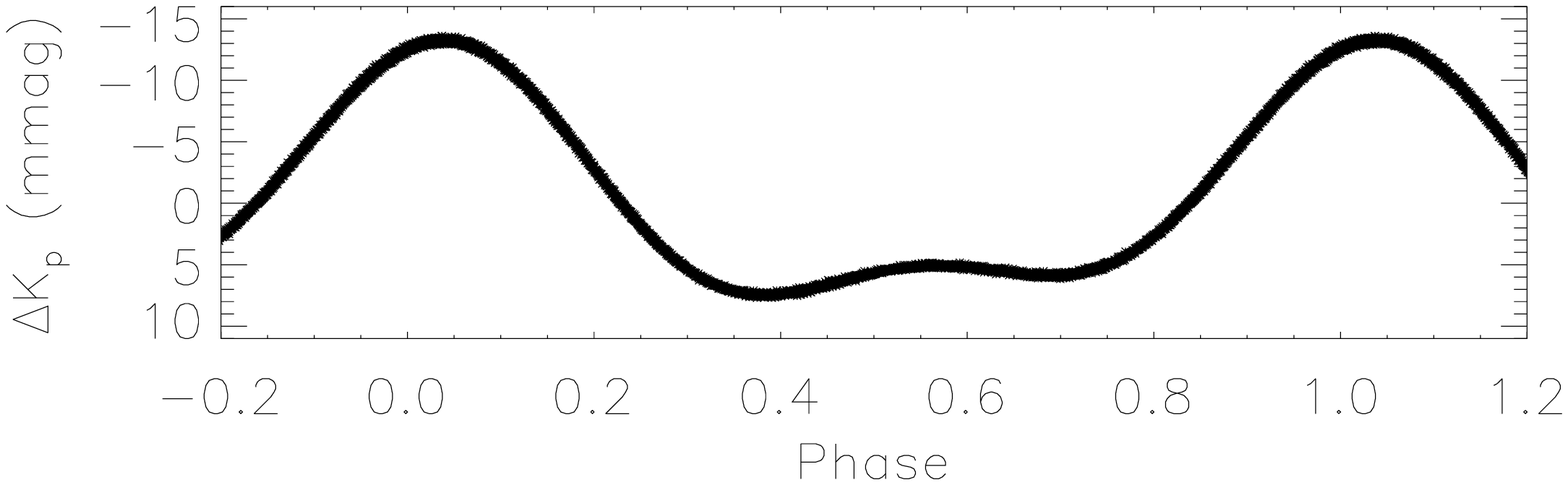}
  \caption{Top panel: The LC data of KIC\,7582608 from Q10. The ordinate is BJD-$245\,0000$. Middle panel: Q$10$\,LC data after high-pass filtering (described in Section\,\ref{sec:kplr-puls}) has removed the rotational modulation signature leaving only the pulsation variation. Only the pulsation envelope is seen here due to the scale. Note that the pulsation maxima coincide with the rotational maxima. Bottom panel: The phase folded light curve.}
  \label{fig:kplr-lc}
\end{figure}

Fig.\,\ref{fig:kplr-lc} shows Q10 of the data. It is clear that the star has a well defined rotation period, mapped out by surface brightness anomalies in two hemispheres. This is typical behaviour of an Ap star which has a strong global magnetic field. To determine the rotation period of the star, we combined all quarters of data, removing any obvious outlying data points and quarter-to-quarter zero-points resulting in a light curve varying about zero magnitude. We then calculated a periodogram which showed two frequencies, the rotation frequency and the harmonic. The principal peak has a frequency of $\nu_{\rm rot} = 0.0488\,920 \pm 0.0000\,003$\,d$^{-1}$, which corresponds to a rotation period of $P_{\rm rot}=20.4532\pm0.0001$\,d. As with the WASP data, the rotational frequency was calculated with P{\sc{eriod}}04, and the errors are the analytical errors. This value is close to that derived from the WASP light curve in Section\,\ref{sec:wasp-rot}, but deviates by $\sim28$\,min. The {\it Kepler} data have a much higher duty cycle than that of the WASP observations, so the daily and seasonal gaps in the WASP data have led to the discrepancies between the two derived rotation periods. Combining the two data sets provides a time string covering 3285\,d from which we calculate a rotation period of $P_{\rm rot}=20.4401\pm0.0005$\,d.

\subsection{The pulsation signature}
\label{sec:kplr-puls}

Due to the barycentric corrections applied to the time stamps of the {\it Kepler} observations, the regular sampling of the data has been broken. \citet{murphy13} have shown that it is possible to directly analyse the true pulsation frequency even if it occurs above the nominal Nyquist frequency of the LC data, rather than analysing a low-frequency alias. This is the case of the pulsation in KIC\,7582608.

\begin{figure}
  \includegraphics[angle=180,width=\linewidth]{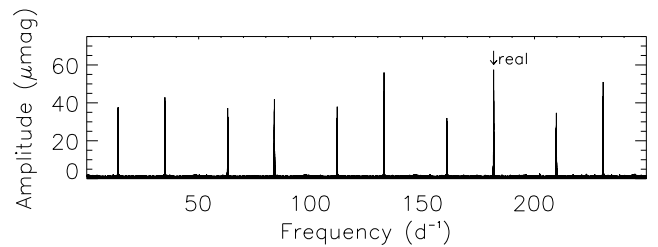}
  \caption{Periodogram of the individually pre-whitened Q00$-$Q17 {\it Kepler} data. The real peak can be identified as the strongest signal.}
  \label{fig:SNA}
\end{figure}

To analyse the high frequency signature, we pre-whiten each quarter of data to the approximate noise level at higher frequencies, thus we remove all peaks with amplitudes greater than $5$\,$\mu$mag below 1\,d$^{-1}$. This procedure removes both instrumental variations at very low frequency, and the rotational spot variations. As these are well-separated in frequency from the 181\,d$^{-1}$ pulsation frequency, the filtering has no effect on our analysis, except to make the noise close to white noise so that the light curve can be examined by eye and so that the least-squares error estimates are realistic (i.e., are not influenced by the low-frequency variations). We initially combined the pre-whitened data and calculated a full periodogram to show the true pulsation in the LC data. Fig.\,\ref{fig:SNA} presents the result, with the highest amplitude peak at $181.7324$\,d$^{-1}$, which agrees with that found from the WASP data in Section\,\ref{sec:WASP-puls}. Note that the amplitude is much lower than that seen in the WASP data as a result of the multiple Nyquist crosses.

\begin{figure}
  \centering
  \includegraphics[angle=180,width=\linewidth]{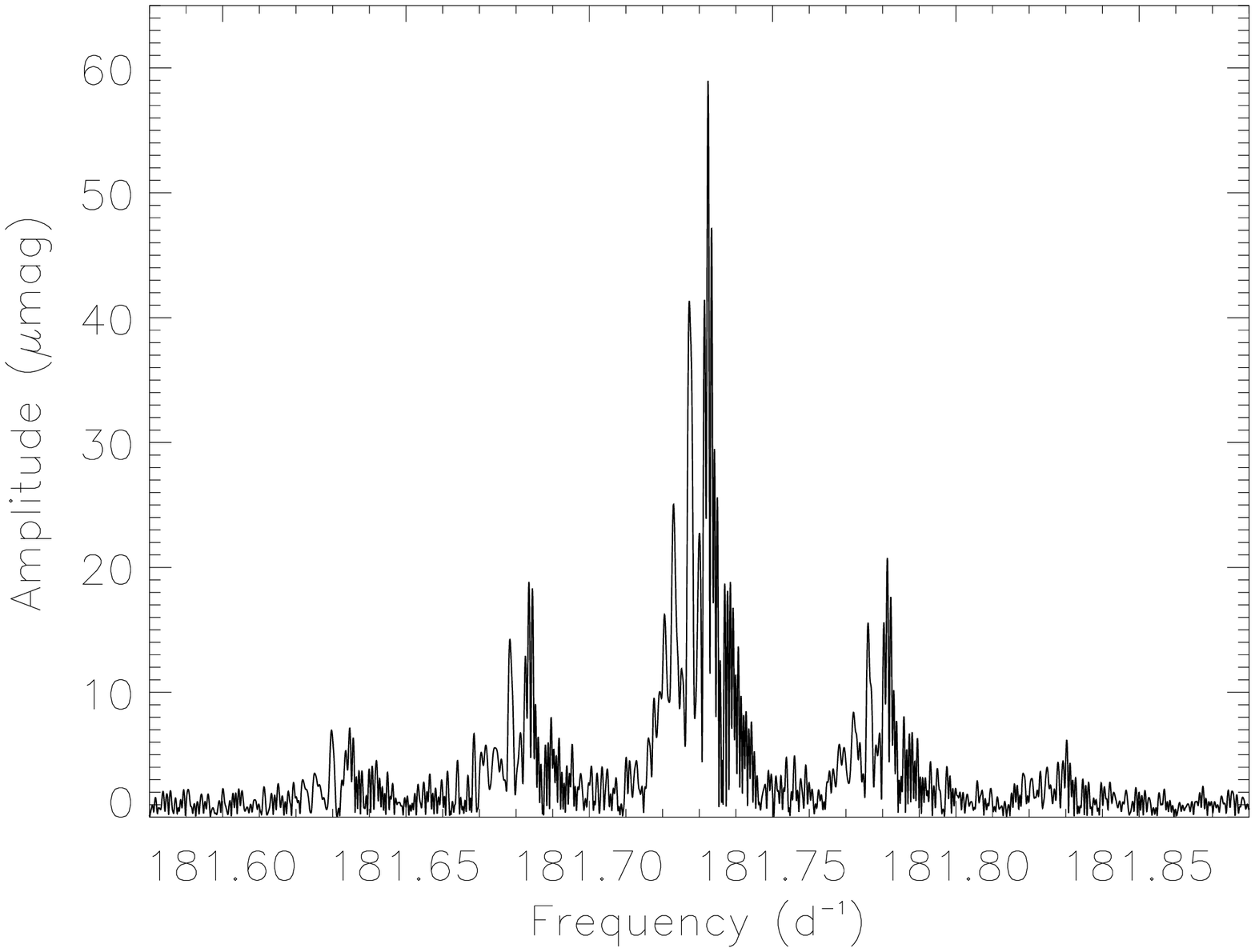}
  \includegraphics[angle=180,width=\linewidth]{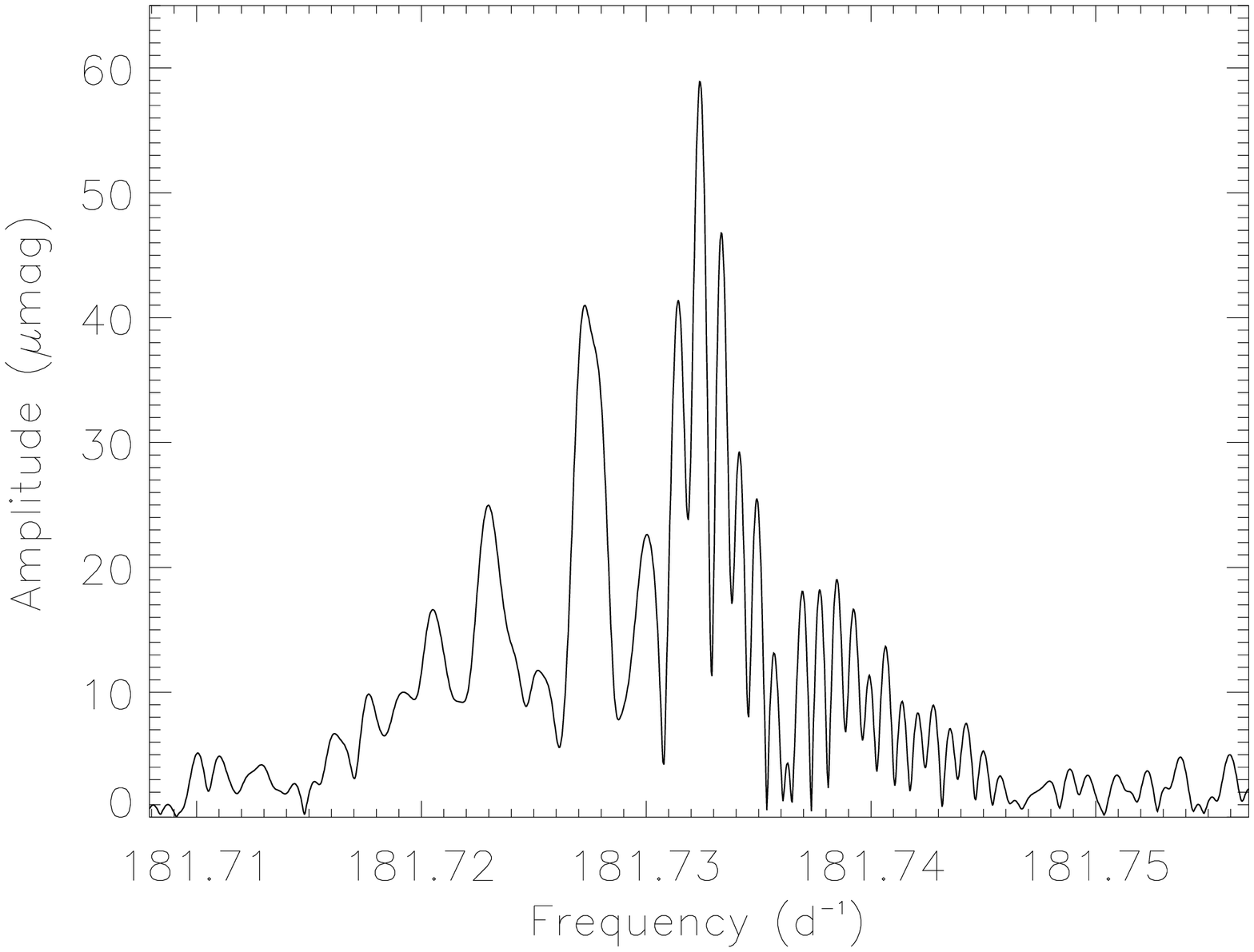}
  \caption{The roAp pulsation frequency seen in the {\it Kepler} Q$00-17$ LC data. Top: An amplitude spectrum showing the multiplicity of closely spaced peak around each of the frequency quintuplet components of the obliquely pulsating mode. Bottom: a higher resolution look at the central peak of the quintuplet. Each of the rotational sidelobes has a similar structure, as expected for oblique pulsation.}
  \label{fig:kplr-high}
\end{figure}

Taking the full data set and extracting just the pulsation range shows, in Fig.\,\ref{fig:kplr-high}, a ``ragged'' multiplet split by the rotation frequency of the star. The central peak is the true pulsation frequency, as confirmed with the WASP data, with the rotationally split side lobes describing the variations in phase and amplitude with the varying aspect over the rotation cycle. A closer look at the principal peak, Fig.\,\ref{fig:kplr-high} bottom, highlights the complex and unstable nature of the pulsation frequency, an observation we address in Section\,\ref{sec:freqvar}.

\begin{figure*}
  \centering
  \begin{minipage}{\textwidth}
    \centering
    \includegraphics[angle=90,width=\textwidth]{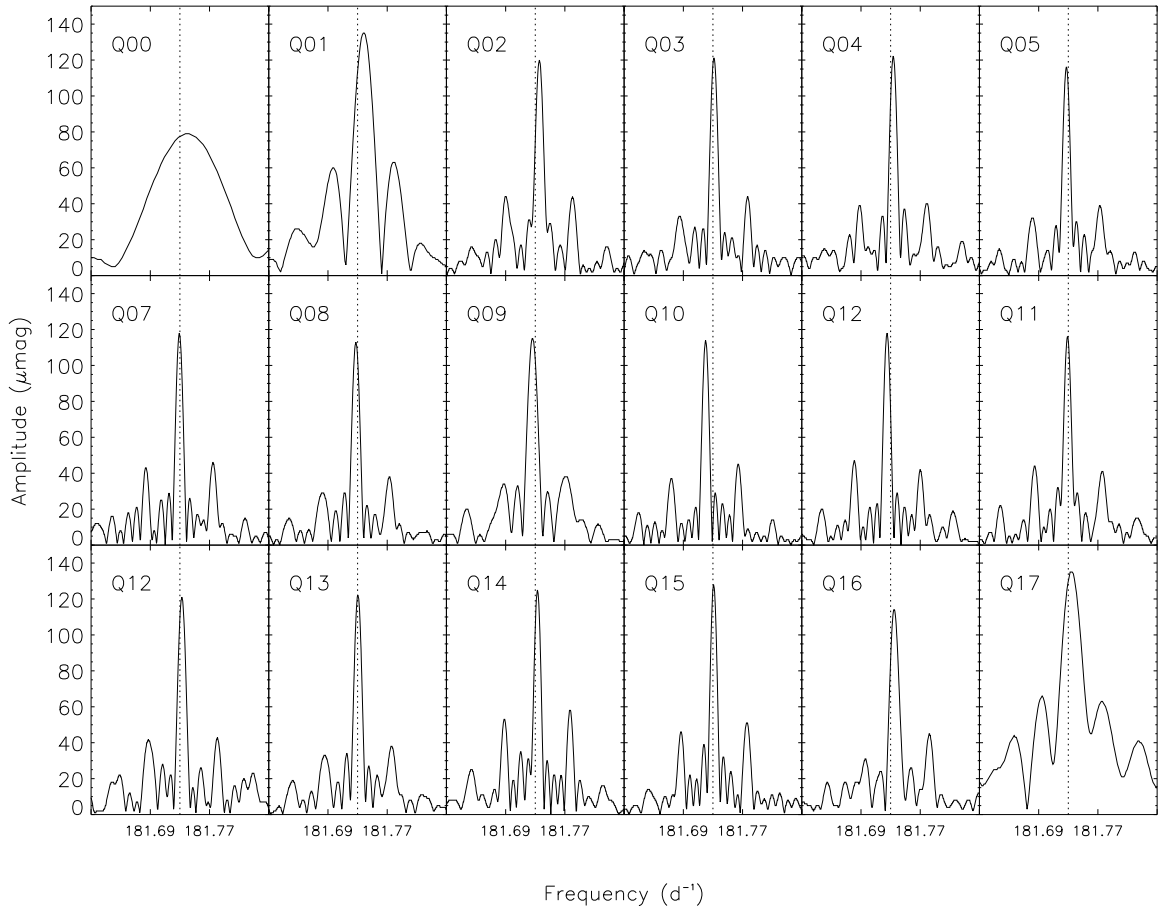}
    \caption{Periodograms of each quarter of {\it Kepler} data centred on the pulsation frequency. The vertical dashed line at 181.73\,d$^{-1}$ is to guide the eye when noting the frequency variability from quarter to quarter, as discussed in Section\,\ref{sec:freqvar}. The values of the pulsation frequency are shown in Table\,\ref{tab:kplr_lls}.}
    \label{fig:montage}
  \end{minipage}
\end{figure*}

We have analysed each quarter individually, with the periodograms shown in Fig.\,\ref{fig:montage} and the results of a linear least-squares fit shown in Table\,\ref{tab:kplr_lls}. To produce the data in Table\,\ref{tab:kplr_lls}, we found the highest amplitude peak in the range $181-182$\,d$^{-1}$ and removed it with a linear least-squares fit. The residuals were then searched for the next highest peak, which was removed in the same way. When we had removed the pulsation signatures from the light curve, we then forced the sidelobes to be separated by exactly the rotation frequency to test the phase relations of the sidelobes in the original data. We find that $\phi_{-1}=\phi_{+1}\neq\phi_0$ suggesting that the mode is a distorted mode.

To test this result, we assume a basic geometry of a pure dipole mode with sidelobes resulting from distortion and rotation alone. In this case, $l=1$ and $m=0$. We use the relationship of \citet{kurtz90} between the amplitude of the central peak and the first sidelobes such as
\begin{equation}
\tan i\tan\beta=\frac{A_{+1}+A_{-1}}{A_0},
\end{equation}
where $i$ is the rotational inclination of the star, $\beta$ is the obliquity of the pulsation axis to the rotation axis, and $A_0$ and $A_{\pm1}$ are the amplitudes of the central and first sidelobes, respectively. Using an average value for the the amplitudes calculated for each quarter of data (Table\,\ref{tab:kplr_lls}), we calculate $\tan i\tan\beta = 0.67\pm0.03$.
This provides us with a relation between $i$ and $\beta$, but not the values individually. For this we require a measure of the $v\sin i$ of the star, and an estimate of the stellar radius. We are unable to measure the $v\sin i$ in Section\,\ref{sec:spec}, however we are able to provide an upper limit of $4$\,km\,s$^{-1}$. Using this upper limit, the radius derived from the \citet{torres10} relationships of of $1.77$\,R$_{\odot}$, and the rotation period of $20.4339$\,d, we can estimate a limit on $i$ of $\sim66^{\circ}$ and on $\beta$ of $\sim17^{\circ}$. From the double wave nature of the light curve shown in Fig.\,\ref{fig:kplr-lc}, it is apparent that $i+\beta > 90^{\circ}$ so that both magnetic poles are seen, a criterion that our calculated values do not fulfil. In fact, no values of $i$ and $\beta$ sum to $>90^{\circ}$ and satisfy $\tan i\tan\beta=0.67$. We therefore calculate $\tan i\tan\beta$ using the assumption that the pulsation is a quadrupole mode such that $l=2$ and $m=0$. Again, from \citet{kurtz90}, we have the relationship between the combined relative amplitudes of the first and second sidelobes which can be used to describe the geometry of the pulsation such as
\begin{equation}
\tan i\tan\beta=4\times\frac{A_{+2}^{(2)}+A_{-2}^{(2)}}{A^{(2)}_{+1}+A^{(2)}_{-1}},
\end{equation}
where $i$ and $\beta$ are as before, and $A^{(2)}_{\pm1}$ and $A^{(2)}_{\pm2}$ are the amplitudes of the first and second sidelobes of the quadrupole mode, respectively. Using an average value for the the amplitudes calculated for each quarter of data (Table\,\ref{tab:kplr_lls}), we calculate $\tan i\tan\beta = 1.44\pm0.14$. Taking the value of $i$ as before, we calculate $\beta$ to be $\sim33^{\circ}$ resulting in $i+\beta > 90^{\circ}$. Such a result suggests that KIC\,7582608 is a quadrupole pulsator.

\subsection{Pulsation variability}
\label{sec:freqvar}

Some roAp stars show highly stable pulsational frequencies, amplitudes and phases over time spans of years. Others have variable frequencies. \citet{kurtz94} and \citet{kurtz97} discussed frequency variability for the roAp star HR\,3831 with ground-based data spanning 16\,y, albeit with large gaps through the years. While they originally suggested that the frequency variability could be cyclic, that was not supported by the later work. Similar frequency variability was reported for another roAp star; \citet{martinez94} discuss HD\,12932 and point out seven further roAp stars for which frequency variability was known at that time.

The question arises as to what causes frequency variability in roAp stars. Is it a change in the pulsation cavity, either because of structural changes in the star, or changes in the magnetic field? Is it externally caused by orbital perturbations of a companion, or companions? Is it some combination of these? 

\citet{balona13} shows an amplitude spectrum for KIC\,10483436 where the largest amplitude pulsation mode has a frequency quintuplet, split by the rotation frequency ($P_{\rm rot} = 4.3$\,d) , caused by oblique pulsation. For the entire {\it Kepler} data set the amplitude spectrum can be described as ``ragged''; that is, the peaks of the quintuplet are composed of many closely spaced peaks in the amplitude spectrum. This is similar in nature to our Fig.\,\ref{fig:kplr-high}. While this is typical for stochastically excited pulsators, such as solar-like stars and red giant stars, stochastic excitation is not likely for roAp stars. Thus the multiplicity of closely spaced peaks that make up each component of the oblique pulsator quintuplet must arise because of frequency, amplitude and/or phase variations over the time span of the data set.

Given the history of non-periodic frequency modulation seen in some roAp stars, we therefore need to study the time dependence of both the frequency of the pulsation mode in KIC\,7582608 and its amplitude. To do this, we used the same data set as in Section\,\ref{sec:kplr-puls}, but split the data into sections with a length of 100 pulsations cycles, or about 0.55\,d. We then fitted $\nu_1 = 181.7324$\,d$^{-1}$, which is the highest peak in Fig.\,\ref{fig:kplr-high}, by linear least-squares to each section of the data, giving 2487 measurements. The function fitted was

\begin{equation}
\Delta m = A \cos (\nu(t - t_0) + \phi)\,.
\label{eq:fn_fit}
\end{equation}

\noindent If the pulsation mode is stable, then we expect to see amplitude and phase modulation only with the rotation frequency 0.0489\,382\,d$^{-1}$ determined in Section\,\ref{sec:wasp-rot} as expected in the oblique pulsator model.

\subsubsection{Amplitude modulation}

\begin{figure}
\begin{center}
\includegraphics[angle=180,width=\linewidth]{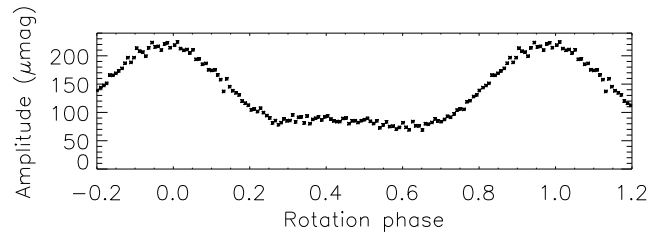}
\caption{Pulsation amplitude as a function of the 20.4339\,d rotation phase, calculated with all quarters of data. The zero point in time, $t_0 = {\rm BJD}\,2453151.8797$, was selected to be the time of maximum rotational brightness of KIC\,7582608, hence we see that maximum pulsation amplitude coincides with the rotational light extremum. }
\label{fig:ampall}
\end{center}
\end{figure}

Fig.\,\ref{fig:ampall} shows the pulsation amplitude as a function of rotational phase for the entire 1460\,d data set. The measurements have been averaged in groups of 20 within narrow phase bins to smooth the curve. It can be seen that there is no change in pulsation amplitude over the entire 1460\,d time span of the data. The only variation seen is the amplitude modulation with rotation caused by the oblique pulsation mode.

This variation is similar to that of another {\it Kepler} roAp star, KIC\,10195926. Figure\,10 of \citet{kurtz11} shows the double wave nature of the amplitude variation with rotation phase, caused by the geometry of the mode and differing aspects of our view. This is also the case here in KIC\,7582608 seen in Fig.\,\ref{fig:kplr-lc} which suggests that both poles may be seen in this star, although with low visibility for the second pole (the small bump seen in Fig.\,\ref{fig:kplr-lc} at rotation phase 0.5).

\subsubsection{Phase and frequency variability}

\begin{figure}
\begin{center}
\includegraphics[width=\linewidth,angle=180]{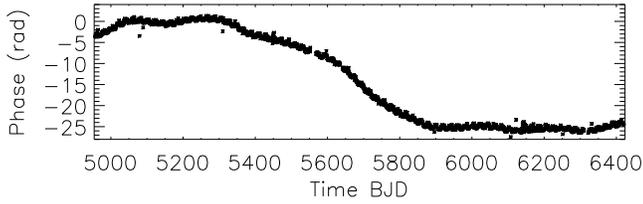}
\end{center}
\caption{The phase variation over the entire observation period. A constant pulsation frequency would led to a straight line in the plot, the variation seen suggest there is a better frequency fit to the data. To form a continuous plot we have added or subtracted $2\pi$ rad to the phase where appropriate.}
\label{fig:phaseall}
\end{figure}

\begin{figure}
\begin{center}
\includegraphics[width=\linewidth,angle=180]{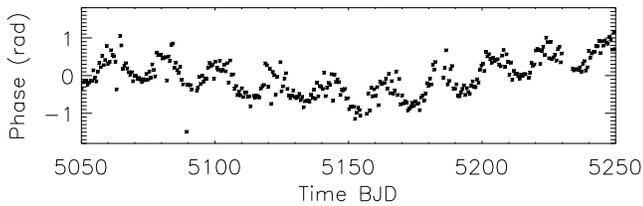}
\end{center}
\caption{This plot shows the phases determined for a section of the 0.55\,d long (100 pulsation cycles) in the time span JD\,$2455048-2455250$.}
\label{fig:phi}
\end{figure}

Unlike the amplitude, the phase varies dramatically over the 4\,y time span. Fig.\,\ref{fig:phaseall} shows the pulsation phase over the entire 1460\,d data set from the same data used for the amplitudes. This is very different from the example of KIC\,10195926 where the pulsation phase is stable over the entire observation period, and shows $\pi$-rad phase reversal at quadrature (figure 10. of \citet{kurtz11}). Selecting a shorter data range of 200\,d, it is possible to see in Fig.\,\ref{fig:phi} that the pulsation phase varies a with rotation phase, as is expected from the oblique pulsator model.

\begin{figure}
\begin{center}
\includegraphics[width=\linewidth,angle=180]{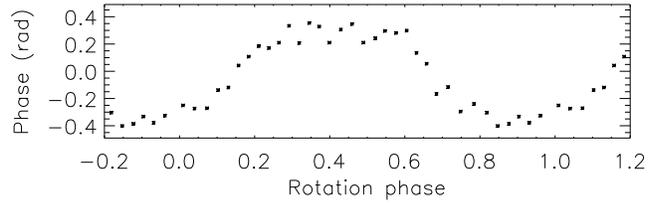}
\end{center}
\caption{This plot shows the phases determined for a section of the 0.55\,d long (100 pulsation cycles) in the time span JD\,$2455048-2455250$. The points are 10-point averages of the 0.55\,d phases in narrow rotational phase bins for smoothing.}
\label{fig:phiphase}
\end{figure}

To examine the rotational variation of the pulsation phase, we have removed the longer term curvature in Fig.\,\ref{fig:phi} and phased the data with the rotation period as seen in Fig.\,\ref{fig:phiphase}. The pulsation phase varies by nearly 1\,rad over the rotation cycle. This is similar to what is seen in other roAp stars that have been studied in enough detail. The best cases to refer to are those of KIC\,10195926 \citep{kurtz11}, HR\,3831 \citep{kurtz97} and HD\,6532 \citep{kurtz96}. For those three stars both pulsation poles are seen, the principal pulsation modes are primarily dipolar, and there is an obvious phase reversal at quadrature. However, outside of that phase reversal, there is smooth pulsation phase variation over the rotation period with an amplitude of order of 1\,rad, just as in KIC\,7582608. Therefore, the only difference for KIC\,7582608 is that the other pulsation pole does not come into sight. This variation of pulsation phase with rotation is a consequence of magnetic and rotational effects on the light curve, along with the geometry of the mode and aspect of our view \citep{bigot11}.

In equation (\ref{eq:fn_fit}) it can be seen that phase and frequency are coupled; a change in one could be interpreted as a change in the other. If we assume that we have a constant phase and only frequency variability, then we may write the light variations as

\begin{equation}
\Delta m = A \cos (\nu(t) (t - t_0) + \phi_0)\, ,
\label{eq:lv}
\end{equation}
where
\begin{equation}
\nu(t) = \nu_0 + \delta\nu(t)
\label{eq:f_def}
\end{equation}
and $\nu_0$ is constant. It is then easy to regroup the terms to give
\begin{equation}
\Delta m = A \cos (\nu_0 (t - t_0) + \phi(t))\, ,
\label{eq:lv2}
\end{equation}
where
\begin{equation}
\phi(t) = \phi_0 + \delta\nu(t) \,(t-t_0)\, .
\label{eq:nu_phi}
\end{equation}
Hence, frequency and phase variability are inextricably intertwined. We therefore interpret the phase variations to be the result of frequency variability.

Fig.\,\ref{fig:phaseall} is the equivalent of a traditional $O-C$ diagram. With a correct constant frequency we expect the phases to follow a straight line in the plot, any linear trend with a slope can mean that another frequency is a better fit. Equations \ref{eq:lv}$-$\ref{eq:nu_phi} allow an easy conversion of phase to frequency. Anywhere in Fig.\,\ref{fig:phaseall} where the trend is non-linear, frequency variations are present.

From the deviations of linearity in Fig.\,\ref{fig:phaseall}, it is possible to see that the pulsation frequency of KIC\,7582608 is strongly variable over the 1460\,d data set. That is the reason for the ``raggedness'' in the amplitude spectrum seen in Fig.\,\ref{fig:kplr-high}.

\begin{figure}
\begin{center}
\includegraphics[width=\linewidth,angle=180]{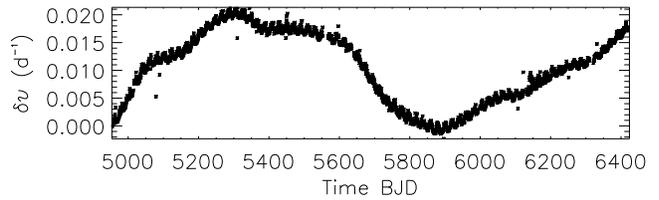}
\end{center}
\caption{The frequency variability, $\delta \nu(t)$, over the 1460\,d data set calculated from Fig.\,\ref{fig:phaseall} with a linear trend removed and equation (\ref{eq:nu_phi}).}
\label{fig:deltaf}
\end{figure}

To examine the frequency variability further, we removed a linear trend from the phases in Fig.\,\ref{fig:phaseall} -- which is equivalent to fitting a different starting frequency initially -- and converted the phases to frequency changes using equation (\ref{eq:nu_phi}). The result is shown in Fig.\,\ref{fig:deltaf}. It can be seen that there is frequency variability on many time scales. Whether the largest variation seen on the length of the data set is a consequence of a binary companion or is intrinsic is uncertain, but clearly much of the frequency variability must be intrinsic to be on so many time scales. This does show why the amplitude spectrum is ``ragged'', although the origin of the frequency variability is not clear.

\subsubsection{Binary interpretation of frequency variation}

One potential cause of frequency variability is external. If the pulsating star is a member of a binary, or multiple system, then the orbital motion causes frequency variability as a result of the Doppler shifts. This is a periodic phenomenon and has been described in detail by \citet{shibahashi12}, who show that the orbital motion results in each pulsation frequency peak being split into a multiplet separated by the orbital frequency. They call their technique for studying binary motion FM, for frequency modulation. The number of components to the multiplet is a function of a parameter $\alpha$, which itself is dependant on the mass of the companion, the orbital period, the pulsation period and the eccentricity; in general, for low eccentricity and low $\alpha$ only a triplet is expected. Fig.\,\ref{fig:kplr-high} is not reminiscent of such a simple pattern. Periodic amplitude modulation also produces frequency multiplets in the amplitude spectrum; oblique pulsation is a good example of this. Again, Fig.\,\ref{fig:kplr-high} is not reminiscent of periodic amplitude modulation.

\begin{figure}
  \includegraphics[angle=180,width=\linewidth]{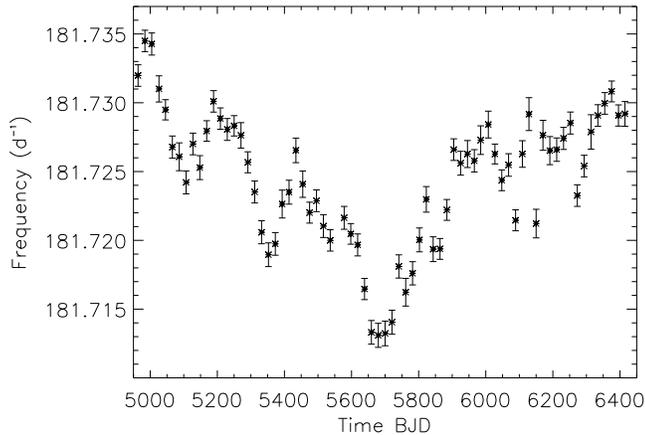}
  \caption{Frequencies calculated for each rotation period using a non-linear least-squares routine.}
  \label{fig:shifts}
\end{figure}

It is evident that in KIC\,7582608 there are frequency variations occurring over many different time scales. As such, Fig.\,\ref{fig:kplr-high} does not fit a simple FM pattern. However, due to the possibility that the large scale frequency variations are due to binary interaction, it is important to investigate this further as, if confirmed, this is an important result for roAp binary systems. To pursue this line of enquiry, we split the pre-whitened Q$00-17$ data into sections of $P_{\rm rot}$ in length, giving 72 individual data sets to analyse. In doing this, we aim to reduce the effects of the shorter time-scale frequency variations which we have shown to be present. This provides us with good temporal coverage of a binary orbit without sacrificing frequency resolution. A periodogram for each rotation cycle was then calculated and fitted with a non-linear least-squares routine to extract the pulsation frequency and corresponding error. In doing this, we extract just the pure frequency variations, allowing the phase to be a free parameter in the fitting procedure. Each individual data set is analysed independently, such that the phases extracted are not of use here. This, therefore, is a different analysis to that presented in Figs.\,\ref{fig:phaseall} and \ref{fig:deltaf}, where we concentrate on phase rather then frequency variations. The results of this analysis are presented in Fig.\,\ref{fig:shifts} and Table\,\ref{tab:Freq-Amp}. It is clear that there are long term frequency variations on approximately the same time period as the observations. The extent of the frequency variations is about 0.02\,d$^{-1}$ (0.25$\mu$Hz).

Assuming that the frequency variability is due to an external factor, i.e. due to orbital motion, it is possible to extract orbital parameters from the frequency variations. Taking the frequency shifts to be caused by Doppler shifts, the frequencies can be converted into radial velocities (RVs) using:
\begin{equation}
V_{\rm rad} = c \left(\frac{\nu_{i}-\nu_{\rm ref}}{\nu_{\rm ref}}\right),
\end{equation}
where $c$ is the speed of light, $\nu_i$ is a given frequency and $\nu_{\rm ref}$ is a reference frequency. As we do not know the intrinsic pulsation frequency, we adopt here $\nu_{\rm ref}$ to be the mean of $\nu$, resulting in relative, rather than absolute, radial velocities. In addition to the {\it Kepler} data, we have included the frequencies of each WASP season of data from Table\,\ref{tab:wasp-nls}, adding a further nine data points, now totalling 81, from which to calculate the binary parameters (Table\,\ref{tab:rv}).

\begin{figure*}
  \begin{minipage}{\textwidth}

    \includegraphics[angle=180,width=85mm]{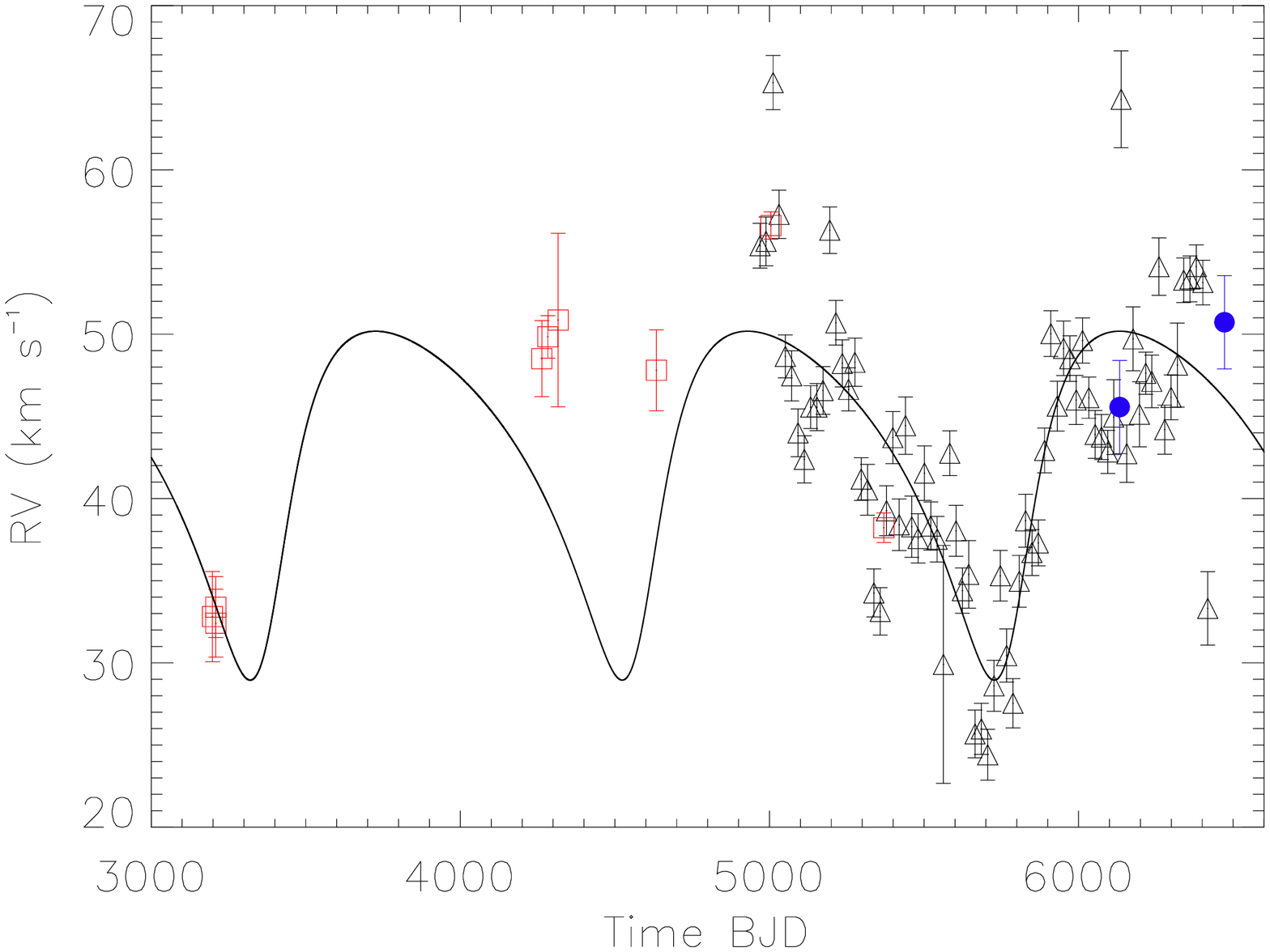}\hfill
    \includegraphics[angle=180,width=85mm]{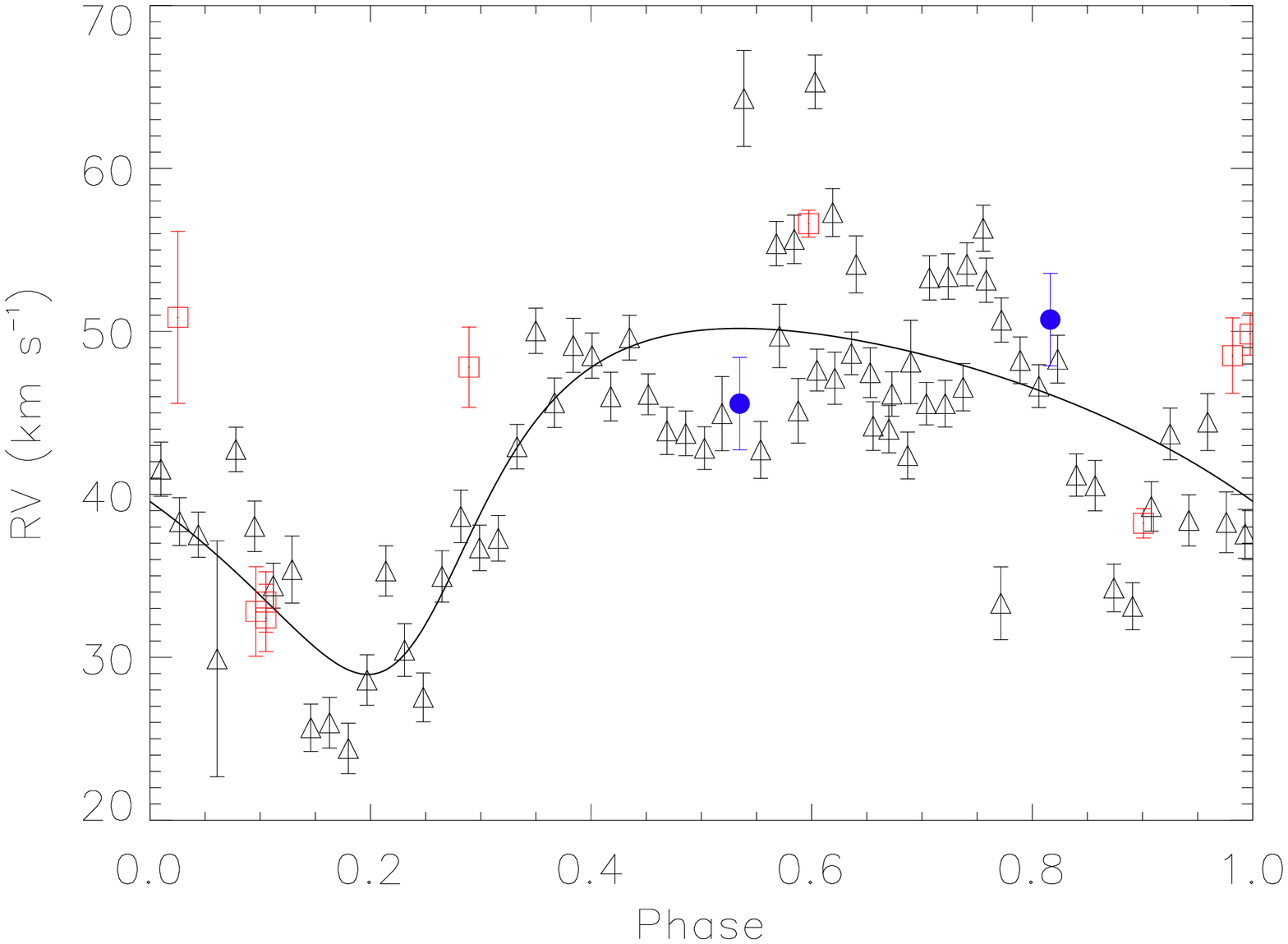}
    \caption{The phase folded eccentric radial velocity curve derived for KIC\,7582608 from photometric data. The black triangles are {\it Kepler} data, the red squares represent WASP data, and the blue dots are derived from the spectra. The line represents the fit from {\sc{jktebop}}. The spectroscopic RVs were not used to determine the fit, but used to determine the absolute value. The fit parameters are shown in Table\,\ref{tab:rv_fit}.}
    \label{fig:rv}
  \end{minipage}
\end{figure*}

To determine the orbital parameters, we pass our RV measurements to {\sc{jktebop}} \citep{southworth13}, testing both circular and eccentric orbits. The best fit the code produces is that of an eccentric system, with an eccentricity of $e=0.39$ and a period of $P_{\rm orb}=1203\pm34$\,d. We have folded the data on the given period and plotted the resultant fit, as shown in Fig.\,\ref{fig:rv}. The full parameters are shown in Table\,\ref{tab:rv_fit}. To determine the absolute values for the RVs, we have included two spectroscopically determined RVs (blue dots in Fig.\,\ref{fig:rv}) from the Lick spectra reported in Section\,\ref{sec:spec}. It must be noted that the spectroscopic RVs are not used in determining the fit, just the absolute values.

\begin{table}
  \centering
  \caption{Radial velocity fit parameters for KIC\,7582608. The mass function was calculated assuming a value of $m_1$ of $1.8$\,M$_\odot$.}
  \label{tab:rv_fit}
  \begin{tabular}{cc}
    \hline
    \multicolumn{1}{c}{Parameter}&
    \multicolumn{1}{c}{Value}\\
    \hline
    $e\cos \omega$ & $-0.301 \pm 0.089$ \\
    $e\sin \omega$ & $-0.247 \pm 0.088$ \\
    $e$ & $0.390 \pm 0.055$ \\
    $\omega$ ($^\circ$)& $219.3 \pm 20.5$\\
    $P_{\rm orb}$ (d) & $1203 \pm 34 $\\
    K$_1$ (km\,s$^{-1}$) & $10.62 \pm 1.13 $\\
    $f(m_1,m_2,\sin i)$ (M$_{\odot}$) & $0.149 \pm 0.016$\\ 
    \hline
  \end{tabular}
\end{table}

The fit is in good agreement with most of the {\it Kepler} and WASP observations. The 2007 and 2008 seasons of the WASP data (between BJD 4200 and 4700 in Fig.\,\ref{fig:rv} left) are perhaps the worst fitting data. The 2008 data are of low quality (see Table\,\ref{tab:wasp-obs}) which may explain the discrepancy. However, we cannot explain the mis-fitting 2007 data, which are of good quality.

Using the mass function presented in Table\,\ref{tab:rv_fit}, we calculate a set of solutions for the secondary mass given a range of binary inclination angles. The results are presented in Table\,\ref{tab:mass_fn}. It is likely that the inclination is close to $90^\circ$ as we detect no signature from a secondary star in the spectra presented in Section\,\ref{sec:spec}, implying a significant luminosity difference.

\begin{table}
  \centering
  \caption{Calculated secondary masses for a given value of the binary inclination.}
  \label{tab:mass_fn}
  \begin{tabular}{cc}
    \hline
    \multicolumn{1}{c}{Inclination $i$ ($^\circ$)}&
    \multicolumn{1}{c}{Secondary mass (M$_\odot$)}\\
    \hline
    90 & 1.07\\
    80 & 1.09\\
    70 & 1.16\\
    60 & 1.30\\
    50 & 1.55\\
    40 & 2.00\\
    \hline
  \end{tabular}
\end{table}

It is clear that KIC\,7582608 is in need of long-term follow-up spectroscopic observations to further monitor the potential RV shifts.

\section{Modelling}
\label{sec:model}

The detection of even a single pulsation frequency in KIC\,$7582608$ may provide additional constraints to the star's global properties.  To investigate this possibility we carried out a linear, non-adiabatic stability analysis of a grid of models covering the theoretical instability strip for roAp stars \citep{cunha02}, restricting the effective temperature to values larger than 7650\,K, taken in intervals of 50\,K. The grid comprises stellar masses between 1.7$\,{\rm M}_{\odot}$ and 2.2$\,{\rm M}_{\odot}$, varying in intervals of 0.05$\,{\rm M}_{\odot}$. The starting point for the grid is a set of evolutionary tracks computed with the code MESA \citep{paxton11,paxton13}, with initial mass fraction of hydrogen and helium of $X=0.70$ and $Y=0.28$, respectively. The effective temperature and luminosities taken from these evolutionary tracks are then used to generate the equilibrium models necessary for the non-adiabatic computations. The analysis follows closely that first presented by \cite{balmforthetal01} and, with some additions, by \cite{cunha13}. An important aspect of the models is that they are composed of two different regions, namely, the equatorial region, where convection proceeds normally, and the polar region, where convection is assumed to be suppressed by the magnetic field.  We refer the reader to the works mentioned above for a detailed description of the models and corresponding physical assumptions.

Following \cite{cunha13}, we have considered, for each set of mass, effective temperature, and luminosity, four different cases, which together cover the main uncertainties in the modelling. The first of these constitutes the standard case, in which the equilibrium model is characterised by the surface helium abundance $Y_{\rm sur}=0.01$ and the minimum optical depth $\tau_{\rm min}=3.5\times10^{-5}$, and the pulsation analysis applies a fully reflective boundary condition at the surface. The other four cases are obtained by swapping these properties, one at the time to: $Y_{\rm sur}=0.1$,  $\tau_{\rm min}=3.5\times10^{-4}$, and a transmissive boundary condition.

\begin{figure}
  \includegraphics[angle=180,width=\linewidth]{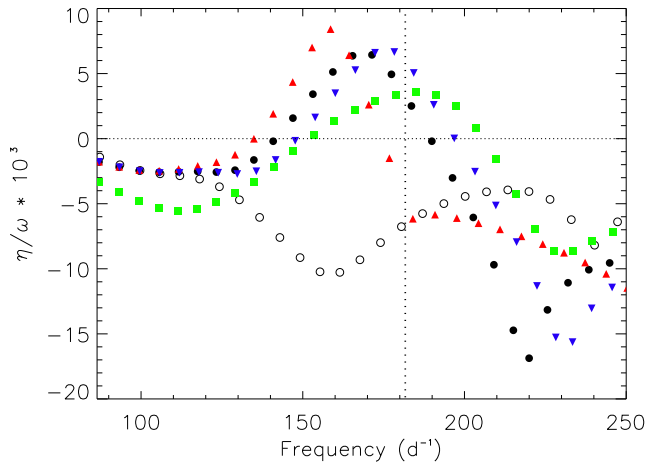}
  \caption{The relative growth rates for the four cases presented in Table\,\ref{tab:models}. The black open circles are for the standard equatorial region (case A, equatorial) and the black filled circles for the standard polar region (case A, polar). The other three cases correspond to modifications to the standard polar region as described in the table: upwards red triangles for case B, down facing blue triangles for case C, and green squares for case D. The vertical dotted line represents the pulsation frequency. The oscillations are stable if the ratio between the imaginary, $\eta$, and the real, $\omega$, parts of the eigenfrequencies is below zero.}
  \label{fig:model}
\end{figure}

Fig.\,\ref{fig:model} shows an example of the results obtained from the stability analysis for a fixed mass, effective temperature and luminosity.  All four cases are shown for the polar region. For the equatorial region, the results of the different cases are very similar and, thus, we present only the standard case. Clearly, the growth rates are negative for all high radial order modes in the equatorial region, indicating pulsation stability when convection takes place normally. This is the usual situation for stars in this region of the HR diagram. In contrast, when convection is suppressed modes at the observed frequency become unstable in three out of the four cases studied for this set of mass, effective temperature and luminosity.

The results of the stability analysis performed on our grid are presented in Fig.\,\ref{fig:HR}. The enclosed region corresponds to the models that show unstable modes at the observed frequency in the polar region of at least one of the four cases studied. Despite the rather weak constraints that exist on the effective temperature and luminosity of KIC\,$7582608$, there  is a clear indication that the results of the stability analysis are consistent with the KIC values, as well as with the effective temperature determinations performed in this work. Thus,  KIC\,$7582608$ seems to be an additional example of the group of stars whose pulsational instability is well explained by the opacity mechanism acting on the hydrogen ionisation region. We note that this is in contrast with a number of cases discussed by \cite{cunha13}, in which clear evidence exists for a disagreement between the observed frequency range and the region of frequencies predicted to be excited by this mechanism, and which led the authors to suggest that an alternative excitation mechanism, possibly connected to the perturbation to the turbulent pressure, must be in place for a subset of roAp stars.

Under the assumption that the opacity mechanism acting in the hydrogen ionisation region is indeed responsible for driving the observed frequency, the results of the stability analysis further constrain the possible values of the luminosity of KIC\,$7582608$ as a function of its effective temperature. An improvement in the determination of the effective temperature and $\log g$ of this star in the future will, thus, lead to a stronger constraint on its luminosity, via the results of the stability analysis. This is because the region of excited modes depends strongly on the radius of the star.

\begin{table}
  \centering
  \caption{Modelling parameters for the cases illustrated in Fig.\,\ref{fig:model},  all computed with  $M=1.8\,{\rm M}_{\odot}$, $T_{\rm eff}=8000$\,K, and $\log g=4.18$.}
  \label{tab:models}
  \begin{tabular}{ccccc}
    \hline
    Case & Model & $Y_{\rm surf}$  & $\tau_{\rm min}$  &Boundary \\
          &        &          &      & condition \\
    \hline
    A & Equatorial & 0.01  & $3.5\times10^{-5}$  &Reflective \\
    A & Polar      & 0.01  & $3.5\times10^{-5}$ & Reflective \\
    B & Polar      & 0.01  & $3.5\times10^{-5}$ & Transmissive \\
    C & Polar      & 0.1   & $3.5\times10^{-5}$  & Reflective \\
    D & Polar       & 0.01 & $3.5\times10^{-4}$  & Reflective \\
    \hline
  \end{tabular}
\end{table}

\section{Conclusion}
\label{conc}

We have analysed two sources of photometric data for the fifth roAp star found in the {\it Kepler} satellite field-of-view. The multi-season WASP data allow us to directly extract the pulsation frequency and amplitude of the star. Although the pulsations are above the Nyquist frequency of the {\it Kepler} LC data, we have also been able to reliably extract the pulsational multiplet frequencies, albeit at greatly diluted amplitudes.

Analysis of the pulsation frequencies has shown that the pulsation mode amplitude is stable, but the frequency is not, leading to ``ragged'' peaks when all data are combined to calculate a periodogram. One possibility for this mode instability is intrinsic variations in the pulsation cavity of the star, leading to slight variations in the frequency over many years.

Another model contributing to the frequency variability we have presented is that KIC\,7582608 is a binary star. We have interpreted some of the frequency variations to be a result of Doppler shifts due to binary motion. Converting these shifts to radial velocities and applying a binary fitting code, we conclude that if the star is indeed in a binary, the orbit must be eccentric, with $e=0.39$, and have a period of about 1200\,d. This result may contradict the suggestion by \citet{scholler12} that magnetic Ap stars become roAp stars if they are not born in close binaries. Of course, the potential secondary object to KIC\,7582608 may have been captured by the Ap star after formation. Further observations are planned for this star to increase the number of spectroscopic RVs to which a binary model can be fit.

Modelling of the pulsation frequency suggests that the pulsation in KIC\,7582608 is driven by the $\kappa$-mechanism acting in the hydrogen ionisation zone, as is the case with most other roAp stars. The mode stability analysis produces results that are consistent with the effective temperature and luminosity of KIC\,7582608 when convection is suppressed in the polar regions by the magnetic field.

\section*{Acknowledgements}

DLH acknowledges financial support from the STFC via the Ph.D. studentship programme. MSC is supported by an Investigador FCT contract funded by FCT/MCTES (Portugal) and POPH/FSE (EC) and by funds from the ERC, under FP7/EC, through the project FP7-SPACE-2012-31284. The WASP project is funded and maintained by Queen's University Belfast, the Universities of Keele, St. Andrews and Leicester, the Open University, the Isaac Newton Group, the Instituto de Astrofisica Canarias, the South African Astronomical Observatory and by the STFC. We thank Luis Balona for providing roAp/noAp data for Fig.\,\ref{fig:HR}, and the referee, Hiromoto Shibahashi, for useful comments and suggestions.

\bibliography{Holdsworth-7582608-refs}

\appendix
\section{Additional Tables}

\begin{table*}
  \scriptsize
  \centering
  \begin{minipage}{\textwidth}
    \centering
    \caption{A linear least-squares fit of the quintuplet. For each Quarter, $t_0$ has been chosen to force the first sidelobes to have equal phase.}
    \label{tab:kplr_lls}
    \begin{tabular}{llcccllccc}
      
      \hline
      {Quarter} & {ID}&   {Frequency}&   {Amplitude}&   {Phase} &  {Quarter} & {ID}&   {Frequency}&   {Amplitude}&   {Phase}\\
      {}&   {}&   {(d$^{-1}$)}&  {($\mu$mag)}&  {(rad)} &   {}&   {}&   {(d$^{-1}$)}&  {($\mu$mag)}&  {(rad)}\\
      
      \hline
     
&       $\nu-2\nu_{\rm rot}$     &$181.6431$ & $19.741 \pm2.485$ & $ 1.659\pm0.126$   & &        $\nu-2\nu_{\rm rot}$	&$181.6268$ & $14.607 \pm1.669$ & $ 0.685\pm0.114$ \\        
&       $\nu-\nu_{\rm rot}$	&$181.6910$ & $40.390 \pm2.537$ & $-1.269\pm0.063$   & & 	$\nu-\nu_{\rm rot}$	&$181.6758$ & $38.438 \pm1.672$ & $ 0.520\pm0.044$ \\        
Q01 &	$\nu$	                &$181.7390$ & $123.843\pm2.544$ & $ 2.259\pm0.021$   &  Q10&	$\nu$	                &$181.7247$ & $113.928\pm1.672$ & $ 1.055\pm0.015$ \\        
&	$\nu+\nu_{\rm rot}$	&$181.7870$ & $36.489 \pm2.538$ & $-1.342\pm0.069$   & & 	$\nu+\nu_{\rm rot}$	&$181.7737$ & $35.002 \pm1.672$ & $ 0.534\pm0.048$ \\        
&	$\nu+2\nu_{\rm rot}$	&$181.8349$ & $11.999 \pm2.487$ & $ 2.168\pm0.207$   & & 	$\nu+2\nu_{\rm rot}$	&$181.8226$ & $15.010 \pm1.669$ & $ 0.674\pm0.111$ \\        
\\                                                                                   
				    	      	     	      	                     
&	$\nu-2\nu_{\rm rot}$	&$181.6377$ & $12.892 \pm1.812$ & $-1.570\pm0.141$   & & 	$\nu-2\nu_{\rm rot}$	&$181.6310$ & $16.549 \pm1.558$ & $ 2.799\pm0.094$ \\        
&	$\nu-\nu_{\rm rot}$	&$181.6868$ & $36.330 \pm1.818$ & $-1.688\pm0.050$   & & 	$\nu-\nu_{\rm rot}$	&$181.6800$ & $42.547 \pm1.559$ & $-0.515\pm0.037$ \\        
Q02&	$\nu$	                &$181.7358$ & $117.870\pm1.817$ & $-1.040\pm0.015$   &  Q11&	$\nu$	                &$181.7290$ & $115.728\pm1.561$ & $-3.072\pm0.014$ \\        
&	$\nu+\nu_{\rm rot}$	&$181.7849$ & $44.590 \pm1.818$ & $-1.666\pm0.041$   & & 	$\nu+\nu_{\rm rot}$	&$181.7781$ & $40.892 \pm1.560$ & $-0.497\pm0.038$ \\        
&	$\nu+2\nu_{\rm rot}$	&$181.8339$ & $14.600 \pm1.812$ & $-1.425\pm0.124$   & & 	$\nu+2\nu_{\rm rot}$	&$181.8271$ & $9.240  \pm1.559$ & $ 2.828\pm0.169$ \\        
\\				    	      	     	      	                     
                                                                                     
&	$\nu-2\nu_{\rm rot}$	&$181.6337$ & $14.135 \pm1.520$ & $ 2.753\pm0.108$   & & 	$\nu-2\nu_{\rm rot}$	&$181.6344$ & $15.754 \pm1.747$ & $-1.083\pm0.111$ \\        
&	$\nu-\nu_{\rm rot}$	&$181.6826$ & $38.813 \pm1.519$ & $-0.438\pm0.039$   & & 	$\nu-\nu_{\rm rot}$	&$181.6838$ & $40.520 \pm1.746$ & $-0.949\pm0.043$ \\        
Q03&	$\nu$	                &$181.7316$ & $122.950\pm1.524$ & $-3.075\pm0.012$   &  Q12&	$\nu$	                &$181.7332$ & $121.494\pm1.748$ & $-0.416\pm0.014$ \\        
&	$\nu+\nu_{\rm rot}$	&$181.7805$ & $43.710 \pm1.523$ & $-0.432\pm0.035$   & & 	$\nu+\nu_{\rm rot}$	&$181.7826$ & $43.475 \pm1.745$ & $-0.917\pm0.040$ \\        
&	$\nu+2\nu_{\rm rot}$	&$181.8295$ & $15.899 \pm1.520$ & $ 2.878\pm0.096$   & & 	$\nu+2\nu_{\rm rot}$	&$181.8319$ & $13.826 \pm1.745$ & $-0.673\pm0.126$ \\        
\\				    	      	     	      	                     
                                                                                     
&	$\nu-2\nu_{\rm rot}$	&$181.6353$ & $15.271 \pm1.533$ & $-2.772\pm0.100$   & & 	$\nu-2\nu_{\rm rot}$	&$181.6329$ & $17.911 \pm1.550$ & $ 2.217\pm0.087$ \\        
&	$\nu-\nu_{\rm rot}$	&$181.6846$ & $39.188 \pm1.536$ & $ 0.172\pm0.039$   & & 	$\nu-\nu_{\rm rot}$	&$181.6817$ & $41.099 \pm1.554$ & $-0.946\pm0.038$ \\        
Q04&	$\nu$	                &$181.7339$ & $123.790\pm1.534$ & $-2.447\pm0.012$   &  Q13&	$\nu$	                &$181.7306$ & $125.120\pm1.553$ & $ 2.776\pm0.012$ \\        
&	$\nu+\nu_{\rm rot}$	&$181.7832$ & $45.875 \pm1.537$ & $ 0.180\pm0.033$   & & 	$\nu+\nu_{\rm rot}$	&$181.7794$ & $39.683 \pm1.553$ & $-0.952\pm0.039$ \\        
&	$\nu+2\nu_{\rm rot}$	&$181.8324$ & $17.410 \pm1.532$ & $-2.740\pm0.088$   & & 	$\nu+2\nu_{\rm rot}$	&$181.8282$ & $11.574 \pm1.551$ & $ 2.600\pm0.134$ \\        
\\				    	      	     	      	                     
                                                                                     
&	$\nu-2\nu_{\rm rot}$	&$181.6312$ & $14.139 \pm1.607$ & $-2.986\pm0.114$   & & 	$\nu-2\nu_{\rm rot}$	&$181.6369$ & $17.362 \pm2.279$ & $-0.798\pm0.131$ \\        
&	$\nu-\nu_{\rm rot}$	&$181.6792$ & $37.914 \pm1.616$ & $-2.887\pm0.043$   & & 	$\nu-\nu_{\rm rot}$	&$181.6850$ & $41.355 \pm2.298$ & $-0.933\pm0.056$ \\        
Q05&	$\nu$	                &$181.7272$ & $117.505\pm1.612$ & $-2.419\pm0.014$   &  Q14&	$\nu$	                &$181.7331$ & $115.532\pm2.306$ & $-0.366\pm0.020$ \\        
&	$\nu+\nu_{\rm rot}$	&$181.7752$ & $36.364 \pm1.615$ & $-2.907\pm0.044$   & & 	$\nu+\nu_{\rm rot}$	&$181.7812$ & $41.746 \pm2.297$ & $-0.940\pm0.055$ \\        
&	$\nu+2\nu_{\rm rot}$	&$181.8231$ & $13.575 \pm1.607$ & $-2.596\pm0.119$   & & 	$\nu+2\nu_{\rm rot}$	&$181.8292$ & $15.442 \pm2.279$ & $-1.038\pm0.148$ \\        
\\				    	      	     	      	                     
                                                                                     
&	$\nu-2\nu_{\rm rot}$	&$181.6312$ & $9.305  \pm1.850$ & $-3.064\pm0.199$   & & 	$\nu-2\nu_{\rm rot}$	&$181.6331$ & $16.573 \pm1.595$ & $-1.985\pm0.096$ \\        
&	$\nu-\nu_{\rm rot}$	&$181.6802$ & $36.056 \pm1.854$ & $-2.765\pm0.051$   & & 	$\nu-\nu_{\rm rot}$	&$181.6821$ & $35.849 \pm1.601$ & $-2.145\pm0.045$ \\        
Q06&	$\nu$	                &$181.7292$ & $113.495\pm1.854$ & $-2.212\pm0.016$   &  Q15&	$\nu$	                &$181.7311$ & $123.619\pm1.604$ & $-1.627\pm0.013$ \\        
&	$\nu+\nu_{\rm rot}$	&$181.7782$ & $41.683 \pm1.854$ & $-2.762\pm0.044$   & & 	$\nu+\nu_{\rm rot}$	&$181.7802$ & $43.531 \pm1.600$ & $-2.152\pm0.037$ \\        
&	$\nu+2\nu_{\rm rot}$	&$181.8272$ & $11.886 \pm1.850$ & $-2.326\pm0.157$   & & 	$\nu+2\nu_{\rm rot}$	&$181.8292$ & $13.050 \pm1.594$ & $-1.965\pm0.122$ \\        
\\				    	      	     	      	                     
                                                                                     
&	$\nu-2\nu_{\rm rot}$	&$181.6301$ & $14.240 \pm1.695$ & $-0.707\pm0.119$   & & 	$\nu-2\nu_{\rm rot}$	&$181.6383$ & $18.673 \pm1.696$ & $-2.639\pm0.091$ \\        
&	$\nu-\nu_{\rm rot}$	&$181.6788$ & $38.421 \pm1.697$ & $-0.800\pm0.044$   & & 	$\nu-\nu_{\rm rot}$	&$181.6871$ & $40.539 \pm1.717$ & $ 0.308\pm0.042$ \\        
Q07&	$\nu$	                &$181.7276$ & $113.775\pm1.699$ & $-0.224\pm0.015$   &  Q16&	$\nu$	                &$181.7359$ & $121.373\pm1.713$ & $-2.267\pm0.014$ \\        
&	$\nu+\nu_{\rm rot}$	&$181.7763$ & $38.983 \pm1.697$ & $-0.804\pm0.044$   & & 	$\nu+\nu_{\rm rot}$	&$181.7846$ & $42.846 \pm1.716$ & $ 0.309\pm0.040$ \\        
&	$\nu+2\nu_{\rm rot}$	&$181.8251$ & $10.953 \pm1.696$ & $-0.516\pm0.155$   & & 	$\nu+2\nu_{\rm rot}$	&$181.8334$ & $11.927 \pm1.697$ & $-2.482\pm0.142$ \\        
\\				    	      	     	      	                     
                                                                                     
&	$\nu-2\nu_{\rm rot}$	&$181.6291$ & $14.525 \pm1.796$ & $ 0.428\pm0.124$   & & 	$\nu-2\nu_{\rm rot}$	&$181.6403$ & $15.675 \pm2.683$ & $ 1.605\pm0.171$ \\        
&	$\nu-\nu_{\rm rot}$	&$181.6777$ & $34.324 \pm1.797$ & $ 0.455\pm0.053$   & & 	$\nu-\nu_{\rm rot}$	&$181.6880$ & $40.562 \pm2.694$ & $-1.123\pm0.067$ \\        
Q08&	$\nu$	                &$181.7263$ & $116.102\pm1.803$ & $ 0.998\pm0.016$   &  Q17&	$\nu$	                &$181.7357$ & $126.786\pm2.709$ & $ 2.686\pm0.021$ \\        
&	$\nu+\nu_{\rm rot}$	&$181.7749$ & $37.724 \pm1.799$ & $ 0.459\pm0.048$   & & 	$\nu+\nu_{\rm rot}$	&$181.7835$ & $42.579 \pm2.699$ & $-1.102\pm0.063$ \\        
&	$\nu+2\nu_{\rm rot}$	&$181.8235$ & $13.191 \pm1.797$ & $ 0.657\pm0.136$   & & 	$\nu+2\nu_{\rm rot}$    	&$181.8312$ & $14.389 \pm2.685$ & $ 2.481\pm0.186$ \\
\\				    	      	     	      	

&	$\nu-2\nu_{\rm rot}$	&$181.6230$ & $12.911 \pm1.581$ & $ 1.770\pm0.123$ \\
&	$\nu-\nu_{\rm rot}$	&$181.6711$ & $36.045 \pm1.588$ & $ 1.652\pm0.044$ \\
Q09&	$\nu$	                &$181.7191$ & $112.235\pm1.585$ & $ 2.196\pm0.014$ \\
&	$\nu+\nu_{\rm rot}$	&$181.7672$ & $40.729 \pm1.587$ & $ 1.635\pm0.039$ \\
&	$\nu+2\nu_{\rm rot}$	&$181.8153$ & $11.504 \pm1.586$ & $ 1.822\pm0.137$ \\
\\				    	      	     	      	

\hline
\end{tabular}
\end{minipage}
\end{table*}
\normalsize

\begin{table*}
  \centering
  \begin{minipage}{\textwidth}
  \centering
  \caption{Frequencies used to produce Fig. \ref{fig:shifts}. The errors are calculated from a non-linear least-squares fit.}
  \label{tab:Freq-Amp}
  \begin{tabular}{cccc}
    \hline
    \multicolumn{1}{c}{Time}&
    \multicolumn{1}{c}{Frequency} &
    \multicolumn{1}{c}{Time}&
    \multicolumn{1}{c}{Frequency}\\
    \multicolumn{1}{c}{BJD}&
    \multicolumn{1}{c}{(d$^{-1}$)}&
    \multicolumn{1}{c}{BJD}&
    \multicolumn{1}{c}{(d$^{-1}$)}\\
    \hline

4963.7671 & $181.7320\pm0.0008$ &   5802.3528 & $181.7200\pm0.0009$  \\
4984.2117 & $181.7345\pm0.0008$ &   5822.8063 & $181.7230\pm0.0009$  \\
5168.2941 & $181.7279\pm0.0007$ &   5843.2594 & $181.7194\pm0.0009$  \\
5188.7478 & $181.7301\pm0.0008$ &   5863.7128 & $181.7194\pm0.0008$  \\
5209.2016 & $181.7288\pm0.0008$ &   5884.1660 & $181.7222\pm0.0008$  \\
5229.6560 & $181.7281\pm0.0008$ &   5904.6193 & $181.7266\pm0.0008$  \\
5250.1101 & $181.7283\pm0.0007$ &   5925.0727 & $181.7256\pm0.0009$  \\
5270.5649 & $181.7276\pm0.0009$ &   5945.5263 & $181.7263\pm0.0010$  \\
5291.0198 & $181.7257\pm0.0008$ &   5965.9804 & $181.7258\pm0.0008$  \\
5311.4745 & $181.7235\pm0.0008$ &   5066.0280 & $181.7268\pm0.0008$  \\
5331.9294 & $181.7206\pm0.0008$ &   5984.8715 & $181.7273\pm0.0010$  \\
5352.3840 & $181.7190\pm0.0009$ &   6007.5842 & $181.7284\pm0.0010$  \\
5004.4821 & $181.7343\pm0.0008$ &   6027.3340 & $181.7263\pm0.0007$  \\
5372.8283 & $181.7197\pm0.0008$ &   6047.7788 & $181.7244\pm0.0008$  \\
5393.2929 & $181.7226\pm0.0010$ &   6067.9677 & $181.7255\pm0.0008$  \\
5413.7468 & $181.7235\pm0.0009$ &   6088.8822 & $181.7215\pm0.0008$  \\
5434.2004 & $181.7265\pm0.0009$ &   6109.1427 & $181.7263\pm0.0010$  \\
5454.6438 & $181.7241\pm0.0010$ &   6128.6775 & $181.7292\pm0.0012$  \\
5475.0869 & $181.7220\pm0.0008$ &   6150.0715 & $181.7212\pm0.0010$  \\
5495.5400 & $181.7229\pm0.0008$ &   6170.5048 & $181.7276\pm0.0011$  \\
5516.0033 & $181.7210\pm0.0008$ &   5086.4811 & $181.7261\pm0.0010$  \\
5536.4567 & $181.7200\pm0.0008$ &   6190.9585 & $181.7265\pm0.0010$  \\
5549.6156 & $181.7340\pm0.0063$ &   6211.4119 & $181.7266\pm0.0009$  \\
5026.0295 & $181.7310\pm0.0010$ &   6231.8650 & $181.7274\pm0.0008$  \\
5577.9772 & $181.7216\pm0.0008$ &   6252.3183 & $181.7285\pm0.0008$  \\
5597.8081 & $181.7205\pm0.0007$ &   6272.7716 & $181.7233\pm0.0008$  \\
5618.2624 & $181.7197\pm0.0008$ &   6293.2248 & $181.7254\pm0.0008$  \\
5638.7170 & $181.7165\pm0.0008$ &   6313.6787 & $181.7279\pm0.0013$  \\
5659.1718 & $181.7133\pm0.0009$ &   6334.1328 & $181.7291\pm0.0008$  \\
5679.6264 & $181.7131\pm0.0009$ &   6354.5869 & $181.7300\pm0.0008$  \\
5700.0815 & $181.7132\pm0.0009$ &   6375.0414 & $181.7308\pm0.0008$  \\
5720.5361 & $181.7140\pm0.0009$ &   5106.9345 & $181.7242\pm0.0008$  \\
5740.9905 & $181.7181\pm0.0008$ &   6395.4963 & $181.7291\pm0.0008$  \\
5761.4448 & $181.7162\pm0.0010$ &   6414.8678 & $181.7292\pm0.0009$  \\
5045.5742 & $181.7295\pm0.0007$ &   5127.3876 & $181.7270\pm0.0008$  \\
5781.8989 & $181.7176\pm0.0008$ &   5147.8410 & $181.7253\pm0.0009$  \\

    \hline
  \end{tabular}
\end{minipage}
\end{table*}

\begin{table*}
  \centering
  \begin{minipage}{\textwidth}
  \centering
  \caption{Radial velocity measurements used to produce Fig. \ref{fig:rv}.}
  \label{tab:rv}
  \begin{tabular}{cccccc}
    \hline
    \multicolumn{1}{c}{Time}&
    \multicolumn{1}{c}{RV} &
    \multicolumn{1}{c}{Orbital}&
    \multicolumn{1}{c}{Time}&
    \multicolumn{1}{c}{RV} &
    \multicolumn{1}{c}{Orbital}\\
    \multicolumn{1}{c}{BJD}&
    \multicolumn{1}{c}{(km\,s$^{-1}$)}&
    \multicolumn{1}{c}{Phase}&
    \multicolumn{1}{c}{BJD}&
    \multicolumn{1}{c}{(km\,s$^{-1}$)}&
    \multicolumn{1}{c}{Phase}\\
    \hline

    \multicolumn{6}{c}{WASP Measurements}\\
\\
 3198.0752 & $33.8740\pm2.7464$ & 0.0961 &  4316.4324 & $51.9240\pm5.2791$ & 0.0254 \\
 3209.0133 & $33.4756\pm2.0667$ & 0.1052 &  4634.0762 & $48.8651\pm2.4607$ & 0.2893 \\
 3209.0175 & $34.4544\pm1.8562$ & 0.1052 &  5004.5471 & $57.6739\pm0.8293$ & 0.5972 \\
 4263.6140 & $49.5729\pm2.3153$ & 0.9815 &  5370.0503 & $39.2848\pm0.9058$ & 0.9009 \\
 4283.0188 & $50.9015\pm1.2920$ & 0.9976 \\ \\

    \multicolumn{6}{c}{\kepler\,Measurements}\\
\\

 4969.5193 & $56.4493\pm1.3612$ & 0.5681 & 5705.8440 & $25.4675\pm1.5437$ & 0.1799 \\
 4988.8706 & $56.7084\pm1.4855$ & 0.5842 & 5726.2986 & $29.6631\pm1.5466$ & 0.1969 \\
 5011.7055 & $66.3719\pm1.6475$ & 0.6031 & 5746.7529 & $36.3605\pm1.5402$ & 0.2139 \\
 5030.8823 & $58.3534\pm1.4691$ & 0.6191 & 5767.2070 & $31.5044\pm1.6158$ & 0.2309 \\
 5051.3364 & $49.7120\pm1.3093$ & 0.6361 & 5787.6611 & $28.6040\pm1.4977$ & 0.2479 \\
 5071.7897 & $48.5287\pm1.5284$ & 0.6531 & 5808.1150 & $36.0218\pm1.5763$ & 0.2649 \\
 5092.2434 & $45.0591\pm1.4496$ & 0.6701 & 5828.5684 & $39.7112\pm1.6092$ & 0.2819 \\
 5112.6968 & $43.4423\pm1.4387$ & 0.6871 & 5849.0217 & $37.7732\pm1.3943$ & 0.2989 \\
 5133.1499 & $46.6247\pm1.2990$ & 0.7040 & 5869.4748 & $38.3586\pm1.3958$ & 0.3159 \\
 5153.6030 & $46.6250\pm1.4314$ & 0.7210 & 5889.9282 & $43.9849\pm1.3632$ & 0.3329 \\
 5173.1674 & $47.6350\pm1.4521$ & 0.7373 & 5910.3813 & $51.0960\pm1.3936$ & 0.3499 \\
 5195.0516 & $57.3907\pm1.4170$ & 0.7555 & 5930.8350 & $46.6841\pm1.5171$ & 0.3669 \\
 5214.9639 & $51.7571\pm1.3571$ & 0.7720 & 5951.2888 & $50.2033\pm1.6614$ & 0.3839 \\
 5235.4182 & $49.2860\pm1.4313$ & 0.7890 & 5971.7429 & $49.5742\pm1.3861$ & 0.4009 \\
 5255.8728 & $47.7015\pm1.3118$ & 0.8060 & 5992.1973 & $47.0606\pm1.4947$ & 0.4179 \\
 5276.3273 & $49.3531\pm1.4662$ & 0.8230 & 6012.6519 & $50.6747\pm1.3699$ & 0.4349 \\
 5296.7822 & $42.2438\pm1.3001$ & 0.8400 & 6033.1067 & $47.1976\pm1.2533$ & 0.4519 \\
 5317.2370 & $41.5897\pm1.5476$ & 0.8570 & 6053.5615 & $44.9607\pm1.4562$ & 0.4689 \\
 5337.6919 & $35.3159\pm1.4623$ & 0.8740 & 6074.0163 & $44.8032\pm1.3711$ & 0.4858 \\
 5358.1467 & $34.1897\pm1.4421$ & 0.8910 & 6094.4709 & $43.8970\pm1.3106$ & 0.5028 \\
 5378.6008 & $40.3191\pm1.5196$ & 0.9080 & 6113.5872 & $46.0190\pm2.2752$ & 0.5187 \\
 5399.0552 & $44.7721\pm1.5833$ & 0.9250 & 6137.5049 & $65.3551\pm2.9457$ & 0.5386 \\
 5419.4988 & $39.4650\pm1.5712$ & 0.9420 & 6155.8340 & $43.7882\pm1.7461$ & 0.5538 \\
 5439.9524 & $45.4938\pm1.7445$ & 0.9590 & 6176.2873 & $50.7768\pm1.9439$ & 0.5708 \\
 5460.4160 & $39.3453\pm1.8627$ & 0.9760 & 6196.7410 & $46.1893\pm1.9796$ & 0.5878 \\
 5480.8693 & $38.6341\pm1.5015$ & 0.9930 & 6217.1943 & $48.6851\pm1.2788$ & 0.6048 \\
 5501.3225 & $42.6083\pm1.6560$ & 0.0100 & 6236.6157 & $48.1910\pm1.5989$ & 0.6210 \\
 5521.7657 & $39.3860\pm1.4660$ & 0.0270 & 6259.9195 & $55.1713\pm1.7476$ & 0.6403 \\
 5542.2293 & $38.5864\pm1.3828$ & 0.0440 & 6278.5542 & $45.2504\pm1.4912$ & 0.6558 \\
 5562.6829 & $30.9665\pm7.2416$ & 0.0610 & 6299.0076 & $47.2231\pm1.3649$ & 0.6728 \\
 5583.1367 & $43.8159\pm1.3629$ & 0.0780 & 6319.4614 & $49.1835\pm2.5602$ & 0.6898 \\
 5603.5908 & $39.0943\pm1.5518$ & 0.0950 & 6339.9153 & $54.3449\pm1.3614$ & 0.7068 \\
 5624.0349 & $35.4591\pm1.3780$ & 0.1119 & 6360.3699 & $54.4262\pm1.3964$ & 0.7238 \\
 5644.4795 & $36.4433\pm2.0544$ & 0.1289 & 6380.7837 & $55.1782\pm1.3183$ & 0.7408 \\
 5664.9344 & $26.7362\pm1.4514$ & 0.1459 & 6401.8716 & $54.2052\pm1.3573$ & 0.7583 \\
 5685.3892 & $27.0453\pm1.5588$ & 0.1629 & 6417.7593 & $34.3778\pm2.2375$ & 0.7715 \\

\\
\multicolumn{6}{c}{Spectroscopic Measurements}\\
\\
 6132.8340 & $ 45.5689\pm2.8384$ & 0.5347 & 6471.7993 & $ 50.7300\pm2.8384$ & 0.8164 \\
 
    \hline
  \end{tabular}
\end{minipage}
\end{table*}

\label{lastpage}

\end{document}